\providecommand{\PRLmode}{all}
\edef\PRLselectedmode{\PRLmode}
\def\PRLalloption{all}
\def\PRLletteroption{letter}
\def\PRLsupplementoption{supplement}
\newif\ifPRLletteronly
\newif\ifPRLsupplementonly
\ifx\PRLselectedmode\PRLletteroption\PRLletteronlytrue\fi
\ifx\PRLselectedmode\PRLsupplementoption\PRLsupplementonlytrue\fi
\ifx\PRLselectedmode\PRLalloption\else
  \ifPRLletteronly\else\ifPRLsupplementonly\else
    \PackageError{PRL manuscript}{Unknown manuscript mode}{Use all, letter, or supplement.}
  \fi\fi
\fi
\ifPRLsupplementonly
  \documentclass[aps,prl,onecolumn,superscriptaddress,nofootinbib,amsmath,amssymb,floatfix]{revtex4-2}
\else
  \documentclass[aps,prl,reprint,superscriptaddress,nofootinbib,amsmath,amssymb,floatfix,longbibliography]{revtex4-2}
\fi

\usepackage{graphicx,bm,booktabs}
\usepackage[T1]{fontenc}
\usepackage[hypertexnames=false,colorlinks=true,linkcolor=blue,citecolor=blue,urlcolor=blue]{hyperref}
\newcommand{\eq}[1]{Eq.~(\ref{#1})}
\newcommand{\Lc}{\mathcal L}
\newcommand{\cR}{\mathcal D_r}
\newcommand{\cT}{\mathcal D_\theta}
\newcommand{\thorn}{\text{\fontencoding{T1}\selectfont\th}}
\newcommand{\papertitle}{Separation of gravitational perturbations of singly spinning Myers--Perry black holes}

\makeatletter
\AtBeginDocument{\immediate\write\@auxout{\string\citation{PRLControl}}}
\makeatother

\ifPRLsupplementonly
\expandafter\def\csname letter@eq:MP\endcsname{1}
\expandafter\def\csname letter@eq:reduced\endcsname{2}
\expandafter\def\csname letter@eq:invariants\endcsname{3}
\expandafter\def\csname letter@eq:symmetries\endcsname{4}
\expandafter\def\csname letter@eq:joint\endcsname{5}
\expandafter\def\csname letter@eq:a0\endcsname{6}
\expandafter\def\csname letter@eq:radI\endcsname{7}
\expandafter\def\csname letter@eq:radII\endcsname{8}
\expandafter\def\csname letter@eq:P\endcsname{9}
\expandafter\def\csname letter@fig:qnm\endcsname{1}
\expandafter\def\csname letter@tab:onset\endcsname{I}
\expandafter\def\csname letter@tab:us\endcsname{II}
\expandafter\def\csname letter@fig:us\endcsname{2}
\newcommand{\lettereq}[1]{\csname letter@#1\endcsname}
\else
\newcommand{\lettereq}[1]{\ref{#1}}
\fi

\ifPRLletteronly
\expandafter\def\csname supplement@eq:master\endcsname{S1}
\expandafter\def\csname supplement@eq:Dr\endcsname{S2}
\expandafter\def\csname supplement@eq:KC\endcsname{S4}
\expandafter\def\csname supplement@eq:K3\endcsname{S5}
\expandafter\def\csname supplement@eq:Fmain\endcsname{S6}
\expandafter\def\csname supplement@eq:Gtriple\endcsname{S7}
\expandafter\def\csname supplement@eq:GPhi\endcsname{S8}
\expandafter\def\csname supplement@eq:vecI\endcsname{S9}
\expandafter\def\csname supplement@eq:vecII\endcsname{S10}
\expandafter\def\csname supplement@eq:JT\endcsname{S24}
\expandafter\def\csname supplement@eq:radcoeffs\endcsname{S27}
\expandafter\def\csname supplement@eq:polynomials\endcsname{S29}
\expandafter\def\csname supplement@eq:phys2\endcsname{S31}
\expandafter\def\csname supplement@sec:operators\endcsname{II}
\expandafter\def\csname supplement@sec:phys\endcsname{VI}
\expandafter\def\csname supplement@sec:psi0deriv\endcsname{III}
\expandafter\def\csname supplement@sec:phi00deriv\endcsname{V}
\expandafter\def\csname supplement@sec:physorigin\endcsname{VII}
\expandafter\def\csname supplement@sec:angular\endcsname{IX}
\expandafter\def\csname supplement@sec:numerics\endcsname{X}
\expandafter\def\csname supplement@eq:chiRK\endcsname{S27a}
\newcommand{\supplementref}[1]{\csname supplement@#1\endcsname}
\else
\newcommand{\supplementref}[1]{\ref{#1}}
\fi

\newcommand{\supplementnumbering}{%
  \setcounter{equation}{0}%
  \setcounter{table}{0}%
  \setcounter{section}{0}%
  \renewcommand{\theequation}{S\arabic{equation}}%
  \renewcommand{\thetable}{S\arabic{table}}%
  \renewcommand{\theHequation}{supp.\arabic{equation}}%
  \renewcommand{\theHtable}{supp.\arabic{table}}%
  \setcounter{secnumdepth}{3}%
}

\begin{document}
\ifPRLsupplementonly
  \title{Supplemental Material for ``\papertitle''}
\else
  \title{\papertitle}
\fi
\author{Jorge E. Santos}
\affiliation{Department of Applied Mathematics and Theoretical Physics, University of Cambridge, Wilberforce Road, Cambridge CB3 0WA, United Kingdom}
\date{October 12, 2026}

\ifPRLsupplementonly
  \supplementnumbering
  \maketitle
  \raggedbottom
  Equation numbers without an S refer to the Letter.
\else
\begin{abstract}
Gravitational perturbations of rotating black holes in higher dimensions generally obey coupled partial differential equations. We show that scalar gravitational perturbations invariant under rotations of the transverse sphere of a singly spinning Myers--Perry black hole separate, despite the obstruction to a decoupled Teukolsky equation. Two commuting, second-order angular operators organise the three gauge-invariant curvature components into joint eigenfunctions. The physical radial problem is a $2\times2$ first-order system, equivalent to a single second-order equation. This gives a separation of the sector containing both the bar-mode and axisymmetric ultraspinning instabilities. Using only ordinary differential equations, we compute the bar-mode quasinormal spectrum in $d=5,6,7$ and find instability onsets at $a/r_+=0.90498$ and $0.83228$ in $d=6,7$. We also compute the growing axisymmetric modes and their onset rotations in $d=6,\ldots,9$. The construction shows how separability can survive the coupling between gravitational polarisations.
\end{abstract}

\maketitle
\raggedbottom
\paragraph*{Introduction.}
The Teukolsky equation for the Kerr black hole is both decoupled and separable~\cite{Teukolsky:1972my,Teukolsky:1973ha}. Decoupling gives a closed equation for one curvature variable. Separation reduces its angular and radial dependence to ordinary differential equations after a Fourier expansion in time and azimuth. In Kerr, this structure provides a direct route to quasinormal modes and mode stability \cite{Whiting:1988vc}. In higher dimensions, the corresponding construction encounters an obstruction: the boost-weight-two Weyl components decouple only on backgrounds admitting a null geodesic congruence with vanishing expansion, rotation and shear~\cite{Durkee:2010qu}. The Myers--Perry (MP) black holes~\cite{Myers:1986un} do not satisfy this condition, although near-horizon geometries of extremal black holes do~\cite{Durkee:2010ea}.

This leaves open whether the coupled gravitational equations can nevertheless separate. An angular harmonic may fix the relative profiles of several curvature components, allowing their radial evolution to be determined by ordinary differential equations even when the components remain coupled. Special sectors of higher-dimensional rotating black holes are known to admit separated equations~\cite{Kodama:2009rq,Murata:2008yx,Lunin:2025gw}, and hidden symmetries give separability for several test fields~\cite{Frolov:2017kze,Lunin:2017drx,Krtous:2018bvk}. A recent development in four dimensions illustrates the distinction between decoupling and separation: the coupled gravito-electromagnetic equations of Kerr--Newman separate through a matrix integrating factor~\cite{Hintz:2026kn}, and their separated spectral problem has been solved numerically~\cite{Destounis:2026kn}.

In this Letter we show that gravitational perturbations of a singly spinning MP black hole separate in the sector invariant under the rotations of its transverse $S^{d-4}$. This sector contains both the bar-mode and ultraspinning instabilities~\cite{Emparan:2003sy,Dias:2009iu,Dias:2010maa,Shibata:2010wz,Dias:2014eua}. Computing its quasinormal spectrum has previously required solving coupled partial differential equations (PDEs) in $(r,\theta)$~\cite{Dias:2014eua}. We construct two commuting angular operators and the physical radial equations in closed form. We then reproduce the bar-mode spectrum~\cite{Dias:2014eua} and compute the growing axisymmetric ultraspinning modes, whose stationary onsets were found in Refs.~\cite{Dias:2009iu,Dias:2010eu}, using ordinary differential equations (ODEs) alone.

\paragraph*{Setup.}
The singly spinning MP metric in $d=n+4$, $n\geq1$, is
\begin{multline}
{\rm d}s^2=-{\rm d}t^2+\frac{r_0^{n+1}}{r^{n-1}\Sigma}\big({\rm d}t-a\sin^2\theta\,{\rm d}\phi\big)^2+\frac{\Sigma}{\Delta}{\rm d}r^2+\Sigma\,{\rm d}\theta^2\\
+(r^2+a^2)\sin^2\theta\,{\rm d}\phi^2+r^2\cos^2\theta\,{\rm d}\Omega_n^2,
\label{eq:MP}
\end{multline}
with $\Sigma=r^2+a^2\cos^2\theta$ and $\Delta=r^2+a^2-r_0^{n+1}r^{1-n}$, where $r_0$ is the horizon radius of the nonrotating black hole with the same mass. The horizon radius $r_+$ is the largest positive root of $\Delta$. For $d\ge6$ a horizon exists for arbitrarily large $a$ at fixed $r_0$; the $d=5$ family instead has $a^2<r_0^2$. Writing ${\rm d}s^2=g_{ab}{\rm d}x^a {\rm d}x^b+Z^2{\rm d}\Omega_n^2$, with $Z=r\cos\theta$, the invariant sector consists of a perturbation of the four-dimensional base metric and of its warp factor. Here $Z$ measures the radius of the transverse sphere at each point of the base; its perturbation describes changes in that radius. This geometric field supplies the third gravitational degree of freedom in addition to the two polarisations of the base metric. For $n=1$ we restrict to the even sector under reflection of the transverse circle; the odd sector decouples. The vacuum equations become
\begin{equation}
R_{ab}=\frac{n}{Z}\nabla_a\nabla_b Z,\qquad Z\Box Z+(n-1)\big[(\nabla Z)^2-1\big]=0,
\label{eq:reduced}
\end{equation}
Thus the curvature of the base is sourced by variations of the transverse sphere, although the full higher-dimensional spacetime is vacuum.
We use the Kinnersley tetrad of the base~\cite{Teukolsky:1972my,Teukolsky:1973ha}, normalised by $\bm l\cdot\bm n=-1$ and $\bm m\cdot\bar{\bm m}=1$, and modes proportional to $e^{-{\rm i}\omega t+{\rm i}m\phi}$. The three curvature variables are
\begin{equation}
\begin{gathered}
\Psi_0=\delta C_{\bm l\bm m\bm l\bm m},\qquad
\bar\Psi_0=\delta C_{\bm l\bar{\bm m}\bm l\bar{\bm m}},\\
\Phi_{00}=\tfrac12\delta R_{\bm l\bm l}.
\end{gathered}\label{eq:invariants}
\end{equation}
Here $C$ and $R_{ab}$ are the Weyl and Ricci tensors of the base. These components vanish on the background and are invariant under linearised diffeomorphisms and tetrad rotations. The bar denotes the $\bm m\leftrightarrow\bar{\bm m}$ partner, rather than complex conjugation of a Fourier mode. We collect them into the triple $\Omega=(\Psi_0,\bar\Psi_0,\Phi_{00})$, which carries the three degrees of freedom: the two base gravitational polarisations and the warp mode.

\paragraph*{Angular separation.}
The aim is to separate the angular dependence from the radial dynamics while retaining the coupling between the three curvature components in $\Omega$. For a scalar field, one chooses an angular harmonic and leaves a single radial amplitude. Here the angular harmonic must have three components: it specifies the angular profiles of $\Psi_0$, $\bar\Psi_0$ and $\Phi_{00}$ together. The gravitational polarisations can therefore mix within one separated mode.

Our central result is a pair of second-order $3\times3$ angular operators, $K_C$ and $K_3$, given explicitly in Eqs.~(\supplementref{eq:KC})--(\supplementref{eq:K3}) of the Supplemental Material (SM)~\cite{SupplementalMaterial}. They are independent of $r$ and $r_0$ and satisfy the exact operator identities
\begin{equation}
[K_C,K_3]=0,\qquad F(K_C,K_3)=0.\label{eq:symmetries}
\end{equation}
The polynomial $F(\lambda,\nu)$ has total degree three: it is quadratic in $\lambda$ and cubic in $\nu$. Its explicit form is given in SM Eq.~(\supplementref{eq:Fmain}). These identities organise the angular problem into joint eigenfunctions,
\begin{equation}
K_C\bm Y_{\lambda\nu}=\lambda\bm Y_{\lambda\nu},\qquad
K_3\bm Y_{\lambda\nu}=\nu\bm Y_{\lambda\nu},\qquad F(\lambda,\nu)=0.
\label{eq:joint}
\end{equation}
For fixed $\omega$ and $m$, $\bm Y_{\lambda\nu}(\theta)$ is one such angular profile, and $\lambda$ and $\nu$ are its separation constants. The relation $F(\lambda,\nu)=0$ means that these constants cannot be chosen independently. Regularity at the angular endpoints selects the allowed harmonics and their eigenvalues (SM Sec.~\supplementref{sec:angular}). Choosing a harmonic fixes the relative angular dependence of the three curvature components; their remaining radial dependence is described below.

$K_C$ extends the angular Teukolsky operator by coupling the base curvature to the warp component. The commuting operator $K_3$ supplies a compatible second angular label, $\nu$, alongside the eigenvalue $\lambda$ of $K_C$. Their explicit form is given in SM Sec.~\supplementref{sec:operators}.

At zero rotation the joint spectrum resolves into the Kodama--Ishibashi (KI) scalar, vector and tensor sectors~\cite{Kodama:2003jz}. These labels refer to harmonics on the full round horizon $S^{n+2}$. Our symmetry restriction instead selects the constant scalar harmonic, $\ell_S=0$, on the transverse $S^n$; it admits modes from all three KI sectors. We denote the KI angular label by $\tilde\ell$, with subscripts S, V and T specifying the sector. For $m=2$, with $C_{\tilde\ell}=\tilde\ell(\tilde\ell+n+1)$,
\begin{equation}
(\lambda,\nu)=\begin{cases}\big(2n+8-C_{\tilde\ell},\,-\tfrac{n}{n+2}C_{\tilde\ell}\big) & \text{scalar},\\ \big(n+8-C_{\tilde\ell},\,0\big) & \text{vector},\\ \big(6-C_{\tilde\ell},\,C_{\tilde\ell}\big) & \text{tensor},\end{cases}
\label{eq:a0}
\end{equation}
At $a\omega=0$, $F$ is a product of three linear factors, one for each KI sector. The condition $F=0$ therefore describes three straight lines in the $(\lambda,\nu)$ plane. For generic nonzero $a\omega$, these become branches of a single irreducible spectral curve. Thus the three KI sectors belong to one coupled angular problem once rotation is included.

\begin{figure}[!tbp]
\includegraphics[width=\columnwidth]{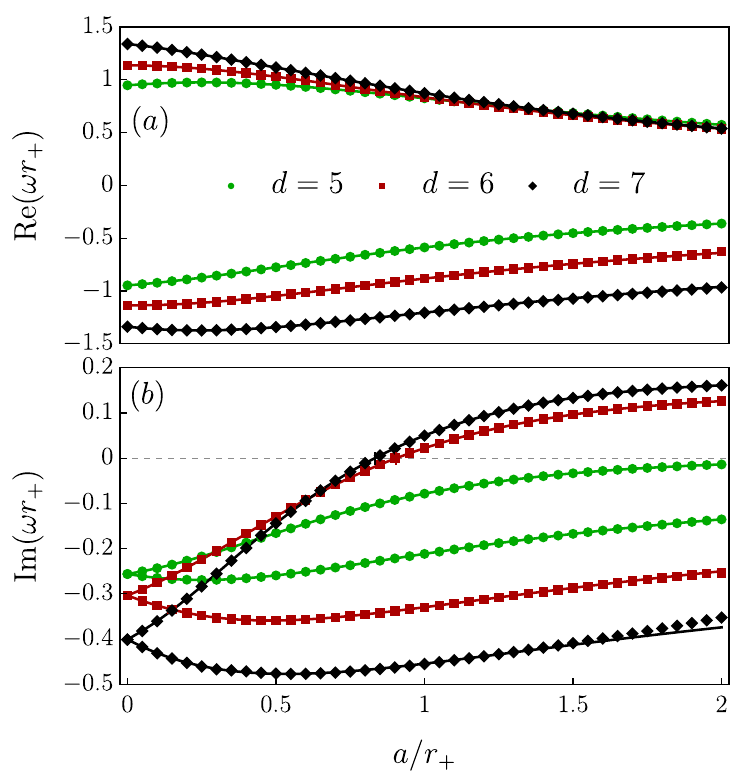}
\caption{Real (a) and imaginary (b) parts of the $\ell_S=0$, $m=2$ quasinormal frequencies connected to the $\tilde\ell_S=2$ KI scalar modes for $d=5,6,7$, as identified by the legend. Solid curves: this work, using the separated angular and radial equations. Filled symbols: the PDE results of Ref.~\cite{Dias:2014eua} (Fig.~17), using the original figure data. Coloured ticks on the dashed zero-growth line in (b) mark the onset of the bar-mode instability.}
\label{fig:qnm}
\end{figure}

\paragraph*{Radial equations.}
The next step is to determine how a chosen angular profile propagates radially. There is an important difference from the vacuum Kerr problem: the warp factor $Z$ of the transverse sphere sources the curvature of the four-dimensional base. The equations for $\Omega$ therefore involve further curvature components. We collect these into a second gauge-invariant triple, $\bm G$, of boost weight zero. Its first two components are rescaled perturbations of the Ricci tensor of the base; the third combines Weyl and Ricci curvature perturbations. Their definitions are SM Eqs.~(\supplementref{eq:Gtriple})--(\supplementref{eq:GPhi}). Both triples arise from the same perturbation of the metric and warp factor, so their radial amplitudes must satisfy additional relations.

The Ricci and Bianchi identities of the base, written in Newman--Penrose form, together with the Hessian--Codazzi identities for $Z$, give the two vector equations in SM Eqs.~(\supplementref{eq:vecI})--(\supplementref{eq:vecII}). The key point is that their radial operators act identically on each component. For a single joint harmonic, the first vector equation fixes the two triples to share the same angular profile:
$$
\Omega=f(r)\bm Y_{\lambda\nu}(\theta),\qquad
\bm G=g(r)\bm Y_{\lambda\nu}(\theta).
$$
In the second vector equation, $K_3$ acting on $\Omega$ becomes multiplication by $\nu$. The angular dependence has now separated: for each harmonic, the unknowns are just the two radial amplitudes $f(r)$ and $g(r)$.

Define the first-order radial operator $D=\partial_r-{\rm i}K/\Delta$, where $K(r)=(r^2+a^2)\omega-am$. With $\cR$ the second-order radial operator in SM Eq.~(\supplementref{eq:Dr}), the separated equations are
\begin{align}
&r^{-\frac n2}\cR\big(r^{\frac n2}f\big)+\Big[\lambda-\tfrac{n(n+6)}{4}\Big]f+2(n+2)\,g=0,\label{eq:radI}\\
&r^{-(n+6)}D\big(r^{n+6}D g\big)+\frac{2(n+3)}{r^2}\,g\nonumber\\
&\qquad=\Bigg\{\frac{\nu}{2r^2}+\frac{{\rm i}n\big[\omega(r^2-a^2)+am\big]}{2r^3}\Bigg\}f.\label{eq:radII}
\end{align}
Equation~(\ref{eq:radI}) is a Teukolsky-type equation for $f$, sourced by the boost-weight-zero amplitude $g$. Equation~(\ref{eq:radII}) transports $g$ along the null direction $\bm l$, with $f$ as its source.

These two equations do not yet give the full physical radial problem. Since each is second order in $r$, one would ordinarily prescribe four initial values, $f$, $f'$, $g$ and $g'$, at a regular radius. Eliminating $g$ at this stage would give a fourth-order equation for $f$. But the two curvature triples come from the same gravitational perturbation, and these four initial values cannot be chosen independently. The remaining Newman--Penrose identities impose the two first-order relations
\begin{equation}
\begin{aligned}
&(\nu+\chi)\,f'+P_f\,f+P_g\,g=0,\\
&(\nu+\chi)\,g'+Q_f\,f+\big[R_K(\nu+\chi)-\chi'-P_f\big]\,g=0.
\end{aligned}
\label{eq:P}
\end{equation}
Here primes denote radial derivatives, and $\chi(r)$ is a scalar coefficient defined in SM Eq.~(\supplementref{eq:chiRK}). The other coefficients $P_f$, $P_g$, $Q_f$ and $R_K$ are also rational functions of $r$ and the mode parameters; their explicit definitions are SM Eqs.~(\supplementref{eq:radcoeffs})--(\supplementref{eq:polynomials}) in Sec.~\supplementref{sec:phys}.

At a regular radius, Eqs.~(\ref{eq:P}) determine $f'$ and $g'$ from $f$ and $g$. Only two initial values remain, giving two independent physical radial solutions for each angular harmonic. The mixture of curvature polarisations is already specified by the chosen harmonic; these two solutions count its radial behaviours. This reduction is consistent with the second-order equations: on the joint spectrum $F(\lambda,\nu)=0$, any solution of Eqs.~(\ref{eq:P}) also satisfies Eqs.~(\ref{eq:radI})--(\ref{eq:radII}). Equivalently, initial data satisfying the physical relations continue to satisfy them under radial evolution.

One can now eliminate $g$ to obtain a single second-order equation for $f$, SM Eq.~(\supplementref{eq:phys2}), with $g$ recovered algebraically. Its extra singular points are apparent: they introduce no additional physical boundary conditions. Thus, although the curvature components remain coupled before separation, each angular harmonic has a radial equation of the same order as the four-dimensional Teukolsky equation. The two physical solutions can be distinguished by their ingoing and outgoing behaviour at the horizon.

A Frobenius analysis at the horizon gives $f\sim(r-r_+)^{-2-{\rm i}\kappa}$ and $f\sim(r-r_+)^{{\rm i}\kappa}$, with $\kappa=K(r_+)/\Delta'(r_+)$, corresponding to ingoing and outgoing solutions, respectively. At infinity, for $\omega\ne0$, the two behaviours are $f_{\rm in}\sim e^{-{\rm i}\omega r}r^{-(n+2)/2}$ and $f_{\rm out}\sim e^{+{\rm i}\omega r}r^{-(n+6)/2}$. With the $e^{-{\rm i}\omega t}$ convention these correspond to ingoing and outgoing waves, respectively. Quasinormal modes are ingoing at the horizon and outgoing at infinity. The construction applies to generic parameters. The exactly stationary axisymmetric case, $\omega=m=0$, is excluded; instability onsets discussed below are obtained by approaching this limit from growing modes.

The SM gives analytic derivations of the vector identities (Secs.~\supplementref{sec:psi0deriv}--\supplementref{sec:phi00deriv}) and the physical relations~(\ref{eq:P}) (Sec.~\supplementref{sec:physorigin}), together with a direct $n=1$ treatment and the numerical methods.

\paragraph*{Bar-mode instability.}
The $m=2$ bar mode describes a nonaxisymmetric deformation of the horizon with a bar-shaped profile in the rotation plane. With our time convention, $\mathrm{Re}\,\omega$ sets the oscillation frequency and $\mathrm{Im}\,\omega>0$ signals exponential growth.
We first recompute the $\ell_S=0$, $m=2$ modes connected to the $\tilde\ell_S=2$ KI scalar mode of Ref.~\cite{Dias:2014eua}. We measure lengths in units of $r_+$ and report the dimensionless quantities $a/r_+$ and $\omega r_+$. At each trial frequency $\omega$, we solve the angular ODE problem for $(\lambda,\nu)$, then the physical radial problem. We vary $\omega$ until the ingoing amplitude at infinity vanishes; the numerical methods are described in SM Secs.~\supplementref{sec:angular}--\supplementref{sec:numerics}.

At $a=0$ the scalar branch has $(\lambda,\nu)=\big(2,-\tfrac{n}{n+2}C_2\big)$. We obtain $\omega r_+=0.947739-0.256094{\rm i}$, $1.136904-0.303576{\rm i}$ and $1.339158-0.400860{\rm i}$ in $d=5,6,7$, respectively, recovering the known KI scalar-mode spectrum~\cite{Cardoso:2003qd,Yoshino:2005ps}. Additional vector and tensor checks are reported in the Supplemental Material.

\begin{table}[!tbp]
\caption{Onset of the bar-mode instability, $\mathrm{Im}\,\omega=0$; $r_0=r_+(1+a^2/r_+^2)^{1/(n+1)}$.}
\label{tab:onset}
\begin{ruledtabular}
\begin{tabular}{lccc}
 & $a_c/r_+$ & $a_c/r_0$ & $\omega_c r_+$\\\hline
$d=6$, this work & $0.90498$ & $0.74136$ & $0.87108$\\
$d=6$, Ref.~\cite{Dias:2014eua} & $0.903\pm0.002$ & $0.740\pm0.001$ & \\
$d=7$, this work & $0.83228$ & $0.72967$ & $0.95111$\\
$d=7$, Ref.~\cite{Dias:2014eua} & $0.833\pm0.002$ & $0.730\pm0.001$ & \\
\end{tabular}
\end{ruledtabular}
\end{table}
Figure~\ref{fig:qnm} compares our separated calculation with the PDE results of Ref.~\cite{Dias:2014eua}. The modes become unstable in $d=6,7$, with the onsets in Table~\ref{tab:onset}; the tracked $d=5$ branch remains damped over the computed range $0\le a/r_+\le2$. The curves agree well, except on the $d=7$, $\mathrm{Re}\,\omega<0$ branch for $a\gtrsim1.6r_+$. There our imaginary part is lower by up to $0.02$, reaching $-0.37398$ at $a=2r_+$. Our frequencies converge to $10^{-7}$ under variations of angular resolution and radial integration parameters. A separate integration of the fourth-order system agrees to $10^{-7}$ on the positive-real-frequency branch and within $2\times10^{-4}$ on the negative branch. For reference, each frequency is computed in a few seconds on a laptop.

\paragraph*{Ultraspinning instability.}
The ultraspinning mode distorts the shape of the rapidly rotating horizon while preserving axial symmetry.
For $m=0$, the angular operators depend on rotation only through $a\omega$. A zero-frequency limit therefore has the angular eigenvalues of $a=0$, making the onset a radial problem at fixed $(\lambda,\nu)$. We find growing modes $\omega={\rm i}\gamma$, $\gamma>0$, for $d\ge6$. Their amplitudes grow as $e^{\gamma t}$, so $\gamma^{-1}$ is the time for the perturbation to grow by a factor of $e$. Their eigenvalues approach the KI tensor branch, $(\lambda,\nu)=(2-C_{\tilde\ell},C_{\tilde\ell})$, with $\tilde\ell_T=2,4$ for the first two unstable harmonics.
\begin{table}[!tbp]
\caption{Onset of the axisymmetric ($m=0$) ultraspinning instabilities, from the limit $\gamma\to0$ of the unstable modes $\omega={\rm i}\gamma$. The last column gives the stationary zero modes of Ref.~\cite{Dias:2010eu} ($\pm0.003$ for $d=6,7$, $\pm0.005$ for $d=8,9$).}
\label{tab:us}
\begin{ruledtabular}
\begin{tabular}{lcccc}
 & $\tilde\ell_T$ & $a_c/r_+$ & $a_c/r_0$ & Ref.~\cite{Dias:2010eu}\\\hline
$d=6$ & 2 & 4.0290 & 1.5598 & 1.572\\
      & 4 & 6.3030 & 1.8320 & 1.849\\
$d=7$ & 2 & 3.0875 & 1.7138 & 1.714\\
      & 4 & 4.6877 & 2.1412 & 2.141\\
$d=8$ & 2 & 2.7037 & 1.7703 & 1.770\\
      & 4 & 4.0158 & 2.2753 & 2.275\\
$d=9$ & 2 & 2.4911 & 1.7924 & 1.792\\
      & 4 & 3.6381 & 2.3369 & 2.337\\
\end{tabular}
\end{ruledtabular}
\end{table}

\begin{figure}[!tbp]
\includegraphics[width=\columnwidth]{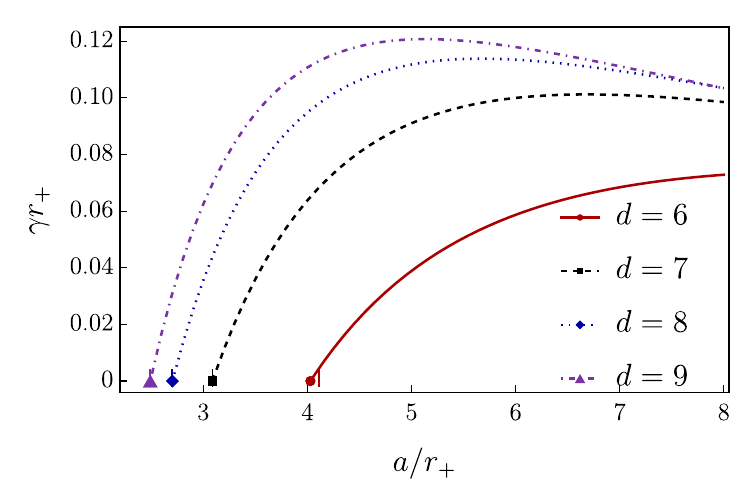}
\caption{Dimensionless growth rate $\gamma r_+$ of the first ultraspinning instability ($m=0$, $\omega={\rm i}\gamma$, connected to $\tilde\ell_T=2$) versus $a/r_+$. Solid, dashed, dotted and dash-dot curves: separated ODE calculation for $d=6,7,8,9$, respectively. Filled symbols: extrapolated onsets. Coloured ticks: stationary zero modes of Ref.~\cite{Dias:2010eu}, converted to $a/r_+$.}
\label{fig:us}
\end{figure}
Figure~\ref{fig:us} shows the growth rates as functions of $a/r_+$. Extrapolating from $\gamma r_+\lesssim3\times10^{-3}$ gives the onsets in Table~\ref{tab:us}. To compare with the original ultraspinning thresholds~\cite{Dias:2010eu}, we convert to $a/r_0=x(1+x^2)^{-1/(n+1)}$, where $x=a/r_+$ and $r_0$ is defined below \eq{eq:MP}. In $d=7,8,9$ this recovers the stationary zero modes to all quoted digits. In $d=6$ the onsets are about $1\%$ lower. Although this exceeds the earlier quoted error estimate, it likely reflects numerical error in the PDE calculation, whose stability and accuracy in $d=6$ were particularly sensitive to the asymptotic fall-off~\cite{Dias:2010eu}. Our separated ODE calculation provides more precise thresholds: linear and quadratic onset extrapolations agree to $10^{-4}$ in $a_c/r_+$ (SM Sec.~\supplementref{sec:numerics}). The sampled growth rates peak near $a/r_+=6.7,5.7,5.1$ in $d=7,8,9$ and then decrease; the $d=6$ rate continues to grow through $a/r_+=8$. Earlier work located stationary thresholds~\cite{Dias:2009iu,Dias:2010eu} and followed the nonlinear evolution~\cite{Figueras:2017zwa}. Our separated ODE analysis also gives the growth rates of the unstable modes.

\paragraph*{Discussion.}
We have separated a coupled gravitational sector of a higher-dimensional rotating black hole that contains both of its known rotational instabilities. The obstruction to a local decoupled equation does not prevent separation: the angular polarisations are selected by a matrix eigenvalue problem, and the radial dynamics reduce to one second-order equation for each joint eigenfunction. This provides a practical route to black hole perturbation spectra and a new structure in the linearised Einstein equations.

It would be useful to identify the covariant origin of the angular operators $K_C$ and $K_3$ in Eqs.~(\ref{eq:symmetries})--(\ref{eq:joint}), and their relation to the principal Killing--Yano tensor~\cite{Frolov:2017kze}. The radial equations also offer a route to studying $d=5$ mode stability, following Whiting's construction~\cite{Whiting:1988vc} and its recent Kerr--Newman extension~\cite{Hintz:2026kn}. Other transverse-sphere harmonics, multiple spins and a cosmological constant are natural extensions.

\paragraph*{Acknowledgements}
I thank all my collaborators for working with me on this topic since I first became interested in it in 2009. In particular, I thank \'Oscar~Dias and Harvey~Reall for patiently listening to the ideas behind my many earlier unsuccessful attempts to achieve decoupling. I am also grateful to them for reading an earlier version of this manuscript and suggesting improvements. The derivations and computations were further checked with the assistance of Claude (Anthropic), under the author’s direction. JES is supported by STFC grant number ST/X000664/1 and partially by Hughes Hall, Cambridge.

\nocite{Dias:2015nua,Geroch:1973am}
\bibliographystyle{apsrev4-2}
\bibliography{references}

\ifPRLletteronly\else
  \onecolumngrid
  \clearpage
  \supplementnumbering
  \section*{Supplemental Material}
  Equation numbers without an S refer to the Letter.
\fi
\fi

\ifPRLletteronly\else
\section{Conventions and reduction}\label{sec:conventions}
We work with the mostly-plus metric~(\lettereq{eq:MP}). Curvature components are $C_{\bm a\bm b\bm c\bm d}=C_{abcd}\,a^ab^bc^cd^d$ and $R_{\bm a\bm b}=R_{ab}a^ab^b$, where $C$ and $R$ are the Weyl and Ricci tensors of the four-dimensional base metric $g_{ab}$. The Kinnersley tetrad is normalised as $\bm l\cdot\bm n=-1$, $\bm m\cdot\bar{\bm m}=1$. For modes proportional to $e^{-{\rm i}\omega t+{\rm i}m\phi}$, we use $K(r)=(r^2+a^2)\omega-am$ and $\zeta=r+{\rm i}a\cos\theta$, with the mode derivative $D=\partial_r-{\rm i}K/\Delta$, throughout this Supplemental Material. The $SO(n+1)$-singlet perturbations are $g_{ab}\to g_{ab}+h_{ab}$ and $Z\to Z+\delta Z$. Off-diagonal components $h_{ai}$ along $S^n$ vanish by symmetry for $n\ge2$; for $n=1$ they form a decoupled sector, odd under $\psi\to-\psi$, which we do not consider. The reduced equations~(\lettereq{eq:reduced}) follow from the reduced action $\int\sqrt{-g}\,Z^n\{R+n(n-1)Z^{-2}[(\nabla Z)^2+1]\}$, up to boundary terms. The linearised system has 11 fields, 4 gauge functions and 3 physical degrees of freedom.

The master equation~(\ref{eq:master}) is obtained by combining the following identities, linearised and projected onto the tetrad: the Bianchi identities of the base, the Ricci identity for the Hessian $H_{ab}=\nabla_a\nabla_bZ$ (written as $\nabla_cH_{ab}-\nabla_bH_{ac}=-R^d{}_{acb}\nabla_dZ$, with $\Phi_{ij}=\frac{n}{2Z}H_{ij}$), and the commutator identity behind Teukolsky's derivation. In the process, $\kappa$, $\sigma$ and $\Psi_1$ are eliminated in favour of the gauge invariant $X_1=2\Phi_{11}\Psi_1-3\Psi_2\Phi_{01}$, which then drops out. Every step was verified in exact arithmetic for symbolic $n$, and separately at $n=1$ (Sec.~\ref{sec:none}).

\section{Operator identities and separation}\label{sec:operators}
This section gives the operators and curvature identities used in Eqs.~(\lettereq{eq:symmetries})--(\lettereq{eq:radII}) of the Letter. The conventions and the quantities $K$ and $\zeta$ are those of Sec.~\ref{sec:conventions}.

\paragraph{The Teukolsky-type operator.}
The Bianchi and Hessian--Codazzi identities give a Teukolsky-type equation with algebraic warp sources. With $\hat\Psi_0=(r\cos\theta)^{n/2}\Psi_0$,
\begin{equation}
(\cR+\cT)\hat\Psi_0=(n+2)(r\cos\theta)^{\frac n2}\Big[\tan^2\theta\,\Phi_{00}-\frac{2\zeta^2}{r^2}\Phi_{02}\Big],
\label{eq:master}
\end{equation}
The second-order radial and angular differential operators are
\begin{align}
\cR&=\frac{1}{\Delta^{2}}\frac{{\rm d}}{{\rm d}r}\Big(\Delta^3\frac{{\rm d}}{{\rm d}r}\Big)+\frac{K^2-2{\rm i}K\Delta'}{\Delta}+8{\rm i}\omega r\nonumber\\
&\quad+\tfrac52\Delta''-3-(a\omega-m)^2-\frac{n(n+2)}{4}\,\frac{a^2-r_0^{n+1}r^{1-n}}{r^2},\label{eq:Dr}\\
\cT&=\frac{1}{\sin\theta}\frac{{\rm d}}{{\rm d}\theta}\Big(\sin\theta\frac{{\rm d}}{{\rm d}\theta}\Big)-\frac{(m+2\cos\theta)^2}{\sin^2\theta}+m^2\nonumber\\
&\quad+a^2\omega^2\cos^2\theta-4a\omega\cos\theta-\frac{n(n+2)}{4\cos^2\theta}.\label{eq:Dth}
\end{align}
For $n=0$ these reduce to the standard four-dimensional Teukolsky operators~\cite{Teukolsky:1972my,Teukolsky:1973ha}. For $n\ge1$, the source terms require the coupled angular problem: \eq{eq:master} is the $\Psi_0$ component of the vector equation~(\ref{eq:vecI}) below. Its $\bar\Psi_0$ partner follows by exchanging $\bm m$ and $\bar{\bm m}$, while the $\Phi_{00}$ component is derived in Sec.~\ref{sec:phi00deriv}.

\paragraph{The commuting angular operators.}
Define $Q=m/\sin\theta-a\omega\sin\theta$, $\Lc_j=\partial_\theta+Q+j\cot\theta$, $\Lc_j^\dagger=\partial_\theta-Q+j\cot\theta$, the Laplacian of $S^{n+2}$ on $SO(n+1)$-invariant functions,
$\Delta_n=(\sin\theta\cos^n\theta)^{-1}\partial_\theta(\sin\theta\cos^n\theta\,\partial_\theta)$,
and the spheroidal potentials
$V_s=(m+s\cos\theta)^2/\sin^2\theta-m^2-a^2\omega^2\cos^2\theta+2sa\omega\cos\theta$.

The commuting operators introduced in Eqs.~(\lettereq{eq:symmetries})--(\lettereq{eq:joint}) are
\begin{align}
(K_C\Omega)_{\Psi_0}&=(\Delta_n-n\tan^2\theta-V_{+2})\Psi_0-(n+2)\tan^2\theta\,\Phi_{00},\nonumber\\
(K_C\Omega)_{\bar\Psi_0}&=(\Delta_n-n\tan^2\theta-V_{-2})\bar\Psi_0-(n+2)\tan^2\theta\,\Phi_{00},\nonumber\\
(K_C\Omega)_{\Phi_{00}}&=\big[\Delta_n-(n+2)\tan^2\theta-V_{0}\big]\Phi_{00}-\tfrac n2\tan^2\theta\,(\Psi_0+\bar\Psi_0),\label{eq:KC}
\end{align}
and
\begin{align}
(K_3\Omega)_{\Psi_0}&=\frac{n(m-a\omega-a\omega\cos^2\theta)}{\cos\theta}\,\Psi_0+\Big[\mathsf c_+^{-1}\Lc^\dagger_{-1}\Lc^\dagger_0\,\mathsf c_+-\tfrac{n(n+2)}{4}\tan^2\theta\Big]\Phi_{00},\nonumber\\
(K_3\Omega)_{\bar\Psi_0}&=-\frac{n(m-a\omega-a\omega\cos^2\theta)}{\cos\theta}\,\bar\Psi_0+\Big[\mathsf c_+^{-1}\Lc_{-1}\Lc_0\,\mathsf c_+-\tfrac{n(n+2)}{4}\tan^2\theta\Big]\Phi_{00},\nonumber\\
(K_3\Omega)_{\Phi_{00}}&=\tfrac{n}{2(n+2)}\,\mathsf c_-^{-1}\big(\Lc_1\Lc_2\,\mathsf c_-\Psi_0+\Lc^\dagger_1\Lc^\dagger_2\,\mathsf c_-\bar\Psi_0\big)-\tfrac{n^2}{8}\tan^2\theta\,(\Psi_0+\bar\Psi_0)-\tfrac{2}{n+2}\Big[\Delta_n-Q^2+\tfrac{(n-1)(n+2)}{\cos^2\theta}\Big]\Phi_{00},\label{eq:K3}
\end{align}
with $\mathsf c_\pm=\cos^{(n\pm2)/2}\theta$. Products act on everything to their right, with the rightmost operator acting first. The off-diagonal entries of $K_3$ contain products of spin-raising and spin-lowering operators. Neither $K_C$ nor $K_3$ directly couples $\Psi_0$ to $\bar\Psi_0$.

\paragraph{The spectral curve.}
The cubic polynomial $F(\lambda,\nu)$ introduced in Eqs.~(\lettereq{eq:symmetries})--(\lettereq{eq:joint}) is, with $\alpha=a\omega$ and $\beta=a\omega-m$,
\begin{align}
nF={}&n\lambda^2\nu-2\lambda\nu^2-(n+2)\nu^3-2n(\beta^2+n+3)\lambda\nu\nonumber\\
&-8n^2\alpha\beta\lambda+2\big[\beta^2-(n+2)(n-1)\big]\nu^2\nonumber\\
&+n\big[\beta^4+2(n+3)\beta^2+4(n^2+2n+3)\alpha\beta+4(n+2)\big]\nu\nonumber\\
&+4n^2\alpha\big[2\beta^3+(n-1)(n+3)\alpha+(n^2+4n+7)\beta\big].
\label{eq:Fmain}
\end{align}
At $a\omega=0$, $nF$ is $\nu(\lambda+\nu-m^2-2)[n\lambda-(n+2)\nu-n(m^2+2n+4)]$, the vector, tensor and scalar branches for every $m$. The three KI sectors are thus continuously connected branches of one spectral problem.

\paragraph{The vector identities.}
Define the boost-weight-zero gauge invariants
\begin{equation}\label{eq:Gtriple}
\bm G=\Big(\frac{\zeta^2}{r^2}\Phi_{02},\ \frac{\bar\zeta^2}{r^2}\Phi_{20},\ G_\Phi\Big),
\end{equation}
where $G_\Phi$ is built from the curvature of the base:
\begin{equation}
2(n+2)r^2G_\Phi=n\bar\zeta^2\,\delta C_{\bm l\bm m\bar{\bm m}\bm n}+n\zeta^2\,\delta C_{\bm l\bar{\bm m}\bm m\bm n}-2\Sigma\,\delta R_{\bm m\bar{\bm m}}+\tfrac16\big[(n+6)r^2-na^2\cos^2\theta\big]\,\delta R .
\label{eq:GPhi}
\end{equation}
Here $R$ is the Ricci scalar of the base, $\Phi_{02}=\frac12\delta R_{\bm m\bm m}$ and $\Phi_{20}=\frac12\delta R_{\bar{\bm m}\bar{\bm m}}$. The background gradient of the combination in \eq{eq:GPhi} vanishes, so $G_\Phi$ is gauge invariant. Then
\begin{align}
&r^{-\frac n2}\cR\big(r^{\frac n2}\Omega\big)+\Big[K_C-\tfrac{n(n+6)}{4}\Big]\Omega+2(n+2)\,\bm G=0,\label{eq:vecI}\\
&r^{-(n+6)}D\big(r^{n+6}D\bm G\big)+\frac{2(n+3)}{r^2}\,\bm G=\frac{K_3\Omega}{2r^2}+\frac{{\rm i}n\big[\omega(r^2-a^2)+am\big]}{2r^3}\,\Omega.\label{eq:vecII}
\end{align}
The $\Psi_0$ component of \eq{eq:vecI} is \eq{eq:master}. The $\Phi_{00}$ components of Eqs.~(\ref{eq:vecI})--(\ref{eq:vecII}) involve exactly the $\Phi_{00}$ rows of $K_C$ and $K_3$, with the same radial operators and the same coupling.
The radial operators act identically on every component of $\Omega$ and $\bm G$. Expanding $\Omega=\sum_k f_k(r)\bm Y_k$ and $\bm G=\sum_k g_k(r)\bm Y_k$ in the joint angular harmonics of Eq.~(\lettereq{eq:joint}) therefore yields Eqs.~(\lettereq{eq:radI})--(\lettereq{eq:radII}) for each harmonic separately.

\section{\texorpdfstring{Derivation of the $\Psi_0$ transport identity}{Derivation of the Weyl transport identity}}\label{sec:psi0deriv}
\paragraph{Ingredients.} We use Newman--Penrose (NP) notation with respect to $g_{\rm NP}=-g$ and the Kinnersley tetrad. Here $\mu$ denotes the NP spin coefficient. On the background $\kappa=\sigma=\lambda=\nu=\epsilon=0$, and
\begin{equation}
\rho=-\frac1{\bar\zeta},\quad \tau=-\frac{{\rm i}a\sin\theta}{\sqrt2\Sigma},\quad \pi=\frac{{\rm i}a\sin\theta}{\sqrt2\bar\zeta^2},\quad \beta=\frac{\cot\theta}{2\sqrt2\zeta},\quad \alpha=\pi-\bar\beta,\quad \mu=-\frac{\Delta}{2\bar\zeta^2\zeta},\quad \gamma=\mu+\frac{\Delta'}{4\Sigma}.
\end{equation}
The remaining mode derivatives are $\boldsymbol\Delta=-\frac{\Delta}{2\Sigma}(\partial_r+{\rm i}K/\Delta)$, $\delta=(\partial_\theta-Q)/(\sqrt2\zeta)$ and $\bar\delta=(\partial_\theta+Q)/(\sqrt2\bar\zeta)$. Derivatives of the warp factor $z=r\cos\theta$ are
\begin{equation}
\frac{Dz}{z}=\frac1r,\qquad \frac{\boldsymbol\Delta z}{z}=-\frac{\Delta}{2r\Sigma},\qquad \frac{\delta z}{z}=-\frac{\tan\theta}{\sqrt2\zeta}.
\end{equation}
Also
\begin{gather}
k\equiv\frac{2\Phi_{11}}{3\Psi_2}=\frac{n\bar\zeta F}{\zeta N},\qquad F=(n+3)r^2+(n-1)a^2\cos^2\theta,\\ N=(n+2)(n+3)r^2-n(n-1)a^2\cos^2\theta-2{\rm i}(n-1)(n+3)ar\cos\theta.
\end{gather}

\paragraph{Codazzi identities.} For $H_{ab}=\nabla_a\nabla_bZ$ define, for tetrad vectors $\bm a,\bm b,\bm c$,
\begin{equation}
\mathcal C(\bm c;\bm a,\bm b)\equiv(\nabla_{\bm c}H)(\bm a,\bm b)-(\nabla_{\bm b}H)(\bm a,\bm c)+R(\nabla Z,\bm a,\bm c,\bm b)=0,
\end{equation}
the Ricci identity for $\nabla Z$, with the Riemann tensor of the base in the NP conventions above, and with $H_{ab}=\frac ZnR_{ab}$ imposed (so that $\Phi_{ij}=\frac n{2Z}H_{ij}$). To first order about MP, the components $(\bm l;\bm l,\bm m)$ and $(\bm l;\bm m,\bm m)$, multiplied by $\frac n{2z}$, read
\begin{align}
&\big[D-2\bar\rho+(1+\tfrac n2)\tfrac{Dz}z\big]\Phi_{01}-\big[\delta+\bar\pi-2\bar\alpha-2\beta+(1+\tfrac n2)\tfrac{\delta z}z\big]\Phi_{00}+\tfrac n2\tfrac{Dz}z\Psi_1-\tfrac n2\tfrac{\bar\delta z}z\Psi_0+2\kappa\Phi_{11}=0,\tag{C1}\\
&\big[D-\bar\rho+(1+\tfrac n2)\tfrac{Dz}z\big]\Phi_{02}-\big[\delta+2\bar\pi-2\beta+(1+\tfrac n2)\tfrac{\delta z}z\big]\Phi_{01}-2\sigma\Phi_{11}-\tfrac n2\tfrac{\boldsymbol\Delta z}z\Psi_0+\tfrac n2\tfrac{\delta z}z\Psi_1=0.\tag{C2}
\end{align}
That is, $({\rm C1})=\frac n{2z}\mathcal C(\bm l;\bm l,\bm m)$ and $({\rm C2})=\frac n{2z}\mathcal C(\bm l;\bm m,\bm m)$ exactly, at linear order; $(\bar{\rm C}1)$, $(\bar{\rm C}2)$ denote the $\bm m\leftrightarrow\bar{\bm m}$ partners in the same normalisation. This normalisation is the one used in every combination below.

\paragraph{Transport equations.} The linearised Bianchi identities (a), (b) of the base, combined with (C1) and (C2), eliminate $\kappa$, $\sigma$ and $\Psi_1$ in favour of $Y\equiv X_1/(3\Psi_2)$. They give two transport equations:
\begin{align}
(D+a_1)Y&=S_1\equiv k(\bar\delta-4\alpha+\pi)\Psi_0-(\delta+\bar\pi-2\bar\alpha-2\beta)\Phi_{00}-(1-k)\Big[\tfrac n2\tfrac{\bar\delta z}{z}\Psi_0+\big(1+\tfrac n2\big)\tfrac{\delta z}{z}\Phi_{00}\Big],\tag{R1}\\
(\delta+b_1)Y&=S_2\equiv k(\boldsymbol\Delta-4\gamma+\mu)\Psi_0-(D-\bar\rho)\Phi_{02}+(1+k)\Big[\tfrac n2\tfrac{\boldsymbol\Delta z}{z}\Psi_0-\big(1+\tfrac n2\big)\tfrac{Dz}{z}\Phi_{02}\Big],\tag{R1$'$}
\end{align}
with $a_1=-2\bar\rho+(1-k)(1+\frac n2)\frac{Dz}{z}$ and $b_1=2\bar\pi-2\beta+(1+k)(1+\frac n2)\frac{\delta z}{z}$. Explicitly, writing each identity as (left-hand side)$\,-\,$(right-hand side),
$$
({\rm R1})=-k\,({\rm a})-(1-k)\,({\rm C1}),\qquad ({\rm R1}')=-k\,({\rm b})+(1+k)\,({\rm C2}),
$$
identically in the linearised NP quantities. These are the same intermediate equations used to obtain Eq.~(\ref{eq:master}).

\paragraph{Step 1: eliminate $Y$.} Apply $(\delta-\bar\alpha-\beta+\bar\pi+b_1)$ to (R1) and $(D-\bar\rho+a_1)$ to (R1$'$), and subtract. With $[\delta,D]=(\bar\alpha+\beta-\bar\pi)D-\bar\rho\,\delta$ on the background, every derivative of $Y$ cancels and the remainder is $c_Y\,Y$, where
\begin{equation}
c_Y=\delta a_1-Db_1-(\bar\alpha+\beta-\bar\pi)\,a_1+\bar\rho\,b_1=0
\end{equation}
identically on the MP background. Hence
\begin{equation}
W\equiv(D-\bar\rho+a_1)S_2-(\delta-\bar\alpha-\beta+\bar\pi+b_1)S_1=0.
\end{equation}

\paragraph{Step 2: remove the Teukolsky part.} The second derivatives of $\Psi_0$ in $W$ are $k(D\boldsymbol\Delta-\delta\bar\delta)\Psi_0$, i.e.\ $k$ times the principal part of the Teukolsky operator. Subtracting $k\,(r\cos\theta)^{-n/2}$ times Eq.~(\ref{eq:master}), written as in its four-dimensional NP form, removes them. Every first derivative of $\Psi_0$ then cancels too, and we find the identity
\begin{equation}
W-k\,(r\cos\theta)^{-n/2}\,\mathcal M=-\frac{\zeta^2}{r^2}\Big\{r^{-(n+6)}D\big(r^{n+6}DG\big)+\frac{2(n+3)}{r^2}G-\frac{(K_3\Omega)_{\Psi_0}}{2r^2}-\frac{{\rm i}n\big[\omega(r^2-a^2)+am\big]}{2r^3}\Psi_0\Big\},
\end{equation}
where $G=\frac{\zeta^2}{r^2}\Phi_{02}$ and $\mathcal M$ is the master equation~(\ref{eq:master}). On shell the left-hand side vanishes, which is the $\Psi_0$ component of Eq.~(\ref{eq:vecII}).

In this derivation the $\Psi_0$ row of $K_3$ is not an input: it is the angular operator left over after $Y$ and the Teukolsky part have been removed. The twice-applied $\delta$ acting on $\Phi_{00}$ produces $\mathsf c_+^{-1}\Lc^\dagger_{-1}\Lc^\dagger_0\mathsf c_+$.

The barred version, with $(\bar\Psi_0,\frac{\bar\zeta^2}{r^2}\Phi_{20})$, follows from the same steps with $\bm m\leftrightarrow\bar{\bm m}$ (also verified exactly, for symbolic parameters).

\section{\texorpdfstring{The $\Phi_{00}$ components of Eqs.~(\ref{eq:vecI})--(\ref{eq:vecII})}{The warp components of the vector identities}}\label{sec:phi00components}
\paragraph{The invariant $G_\Phi$.} The boost-weight-zero, spin-zero invariant in Eqs.~(\ref{eq:vecI})--(\ref{eq:vecII}) is
\begin{equation}
2(n+2)r^2G_\Phi=n\bar\zeta^2\delta\Psi_2+n\zeta^2\delta\bar\Psi_2-2\Sigma\,\delta R_{\bm m\bar{\bm m}}+\frac{n(n-1)\big[na^2x^2-(n+6)r^2\big]}{6r^2x^2}\,\delta\Big[(\nabla Z)^2-1+\frac{r_0^{n+1}Z^2}{r^{n+1}\Sigma}\Big],
\end{equation}
where $x=\cos\theta$ and $r_0^{n+1}r^{-n-1}/\Sigma$ in the last bracket is the fixed background function (the bracket vanishes on the background). It emerges as an output of the derivation in Sec.~\ref{sec:phi00deriv}. The bracket equals $Z^2\delta U$, with $U\equiv[(\nabla Z)^2-1]/Z^2$ and $U=-r_0^{n+1}r^{-n-1}/\Sigma$ on the background, so this form is regular for every $n$. On shell, Eq.~(\lettereq{eq:reduced}) gives $R=-n(n-1)U$. For $n>1$ the last term is therefore $\frac16[(n+6)r^2-na^2x^2]\,\delta R$, which is Eq.~(\ref{eq:GPhi}); for $n=1$ it vanishes, as does $\delta R$. Its gauge invariance is manifest in the form of Eq.~(\ref{eq:GPpsi}) of Sec.~\ref{sec:physorigin}: $G_\Phi$ is a combination of the warp-corrected invariants $\psi_2$, $\bar\psi_2$ and $\varphi$. It was also checked directly, symbolically, for arbitrary $\Delta(r)$.

\paragraph{The identities.} Written out, the $\Phi_{00}$ components are
\begin{align}
&r^{-\frac n2}\mathcal D_r\big(r^{\frac n2}\Phi_{00}\big)+\Big[\Delta_n-(n+2)\tan^2\theta-V_0-\tfrac{n(n+6)}4\Big]\Phi_{00}-\tfrac n2\tan^2\theta\,(\Psi_0+\bar\Psi_0)+2(n+2)G_\Phi=0,\\
&r^{-(n+6)}D\big(r^{n+6}DG_\Phi\big)+\frac{2(n+3)}{r^2}G_\Phi=\frac{(K_3\Omega)_{\Phi_{00}}}{2r^2}+\frac{{\rm i}n\big[\omega(r^2-a^2)+am\big]}{2r^3}\Phi_{00}.
\end{align}
Both follow from the derivation in Sec.~\ref{sec:phi00deriv}; the coupling in the second is the same as in the $\Psi_0$ component. $G_\Phi$ is invariant under $\bm m\leftrightarrow\bar{\bm m}$, consistent with the symmetric $\Phi_{00}$ rows of $K_C$ and $K_3$.

\paragraph{Consequences.} (i) The $\Phi_{00}$ rows of $K_C$ and $K_3$ are not separate inputs. They are the angular operators in the $\Phi_{00}$ components of identities whose radial parts are fixed by the $\Psi_0$ component. (ii) Separation needs no symmetry argument. Expanding $\Omega=\sum f_k(r)\bm Y_k$ and $\bm G=\sum g_k(r)\bm Y_k$ in joint angular harmonics, Eqs.~(\ref{eq:vecI})--(\ref{eq:vecII}) hold componentwise in $k$ because their radial operators are scalar. (iii) Eliminating $\bm G$ gives the fourth-order vector system $\big[\mathcal T_r(\mathcal R+K_C)+(n+2)r^{-2}K_3+2(n+2)c(r)\big]\Omega=0$, where $\mathcal R=r^{-n/2}\mathcal D_rr^{n/2}-\frac{n(n+6)}4$, $\mathcal T_r=r^{-(n+6)}Dr^{n+6}D+\frac{2(n+3)}{r^2}$ and $c=\frac{{\rm i}n[\omega(r^2-a^2)+am]}{2r^3}$. Since $[K_C,K_3]=0$ and both commute with radial operators, $K_C$ and $K_3$ are manifestly symmetries of this system. Its physical solutions satisfy, in addition, the two first-order relations of Sec.~\ref{sec:phys}.

\section{Analytic derivation of the warp components}\label{sec:phi00deriv}
\paragraph{Method.} We use the exact NP equations of the base: Ricci identities, commutators, Bianchi and contracted Bianchi identities. We also use the Hessian--Codazzi identity $\nabla_cH_{ab}-\nabla_bH_{ac}=-R(\nabla Z,e_a,e_c,e_b)$ for $H_{ab}=\nabla_a\nabla_bZ$, the field equation $H_{ab}=\frac Zn R_{ab}$ and the trace equation $E_Z$. Conventions are those of Sec.~\ref{sec:psi0deriv}, with $\Psi_2\equiv C_{\bm l\bm m\bar{\bm m}\bm n}$, $\bar\Psi_2\equiv C_{\bm l\bar{\bm m}\bm m\bm n}$ and $\Lambda=R/24$.

Each identity below is exact. It is linearised about MP automatically, keeping every perturbed spin coefficient, the perturbation of $Z$ and the perturbation of the tetrad. None of these survive in the final results. The combinations were found by exact linear algebra over the space of NP identities and are unique within the sets used. The final identities were verified in exact arithmetic, for symbolic background and mode parameters, with all linearised NP quantities (spin coefficients, curvature components, tetrad perturbation, $\delta Z$) treated as independent functions of $(r,\theta)$: every coefficient vanishes identically. 

\paragraph{Ingredients.} We use the equations labelled (C1), (R1) and their $\bm m\leftrightarrow\bar{\bm m}$ partners from Sec.~\ref{sec:psi0deriv}, together with the identities
\begin{align}
\bar\delta\Psi_0-D\Psi_1+D\Phi_{01}-\delta\Phi_{00}&=\dots,\tag{a}\label{eq:ingredienta}\\
D\Psi_2-\bar\delta\Psi_1+\boldsymbol\Delta\Phi_{00}-\bar\delta\Phi_{01}+2D\Lambda&=\dots,\tag{e}\label{eq:ingrediente}\\
D\Phi_{11}-\delta\Phi_{10}-\bar\delta\Phi_{01}+\boldsymbol\Delta\Phi_{00}+3D\Lambda&=\dots,\tag{CB}\label{eq:ingredientCB}\\
D\bar\rho-\delta\bar\kappa&=\dots+\Phi_{00},\tag{\(\bar{\rm R}\)}\label{eq:ingredientRbar}
\end{align}
where the first two are Bianchi identities and the latter two are the contracted Bianchi identity and the Ricci identity, respectively. To these we add two equations that follow from them on shell.
\begin{itemize}
\item[$\bullet$] The singlet analogue of Durkee--Reall's equation (5.3) for the sphere components of $\Omega_{ij}$. On shell $\Phi_{11}+3\Lambda=\frac{n}{2Z}H_{m\bar m}$. Eliminating $DH_{m\bar m}$ from (CB) with the Codazzi identity on $(\bm l;\bm m,\bar{\bm m})$ gives, up to terms quadratic in quantities that vanish on the background,
\begin{multline}
\big(\boldsymbol\Delta-2\gamma-2\bar\gamma+\mu+\tfrac n2\tfrac{\boldsymbol\Delta Z}{Z}\big)\Phi_{00}-\big(\delta-2\bar\alpha-2\tau+\tfrac n2\tfrac{\delta Z}{Z}\big)\Phi_{10}+\big(2\bar\tau+\tfrac{\bar\delta Z}{Z}\big)\Phi_{01}\\
+\tfrac n2\tfrac{\delta Z}{Z}\bar\Psi_1-2\bar\rho\,\Phi_{11}-\tfrac{DZ}{Z}\Big[\tfrac n2\bar\Psi_2+\Phi_{11}+(n+3)\Lambda\Big]=0 .
\tag{E}\label{eq:ingredientE}
\end{multline}
It contains no derivatives of boost-weight-zero quantities.
\item[$\bullet$] The trace of the field equation together with $E_Z$ gives $\Lambda=-\frac{n(n-1)}{24Z^2}[(\nabla Z)^2-1]$. Differentiating along $\bm l$, with $D(\nabla Z)^2=2H(\bm l,\nabla Z)$:
\begin{equation}
D\Lambda+\frac{n-1}{12Z}\,R(\bm l,\nabla Z)+2\frac{DZ}{Z}\Lambda=0 .
\tag{DL}\label{eq:ingredientDL}
\end{equation}
\end{itemize}

\paragraph{The $\Phi_{00}$ component of Eq.~(\ref{eq:vecI}).} Let
$$
W\equiv D(E)+w_1(E)+\Big(\delta+\tfrac{\delta Z}{Z}+2\beta+4r\bar\rho\tau\Big)(\bar{\rm C}1)+\tfrac n2\tfrac{\delta Z}{Z}(\bar{\rm a})+\tfrac n2\tfrac{DZ}{Z}(\bar{\rm e})+\tfrac{DZ}{Z}({\rm CB})+2\Phi_{11}(\bar{\rm R}),\qquad w_1=\frac{4r^2-{\rm i}arx-a^2x^2}{r\Sigma},
$$
with $x=\cos\theta$, and with $(\bar{\rm C}1)=\frac n{2Z}\mathcal C(\bm l;\bm l,\bar{\bm m})$ normalised as in Eq.~(C1). Then, identically,
\begin{equation}
\frac{\rho W-\bar\rho\bar W}{\rho-\bar\rho}=-\frac{\rho\bar\rho}{2}\Big[r^{-\frac n2}\mathcal D_r\big(r^{\frac n2}\Phi_{00}\big)+(K_C\Omega)_{\Phi_{00}}-\tfrac{n(n+6)}4\Phi_{00}+2(n+2)G_\Phi\Big].
\end{equation}
The structure parallels Teukolsky's derivation:
\begin{itemize}
\item $D(E)$ and $\delta(\bar{\rm C}1)$ remove the second derivatives of $\Phi_{10}$ through $[\delta,D]$;
\item $(\bar{\rm a})$ and $(\bar{\rm e})$ remove $\delta\bar\Psi_0$, $D\bar\Psi_1$ and $D\bar\Psi_2$;
\item the Ricci identity turns $D\delta\bar\rho-\delta\bar\kappa$ into $\Phi_{00}$, as $D\sigma-\delta\kappa$ turns into $\Psi_0$ for Kerr.
\end{itemize}
The twist $\rho-\bar\rho$ in the denominator is harmless, because the numerator vanishes with it. The $\Phi_{00}$ row of $K_C$, the couplings $-\frac n2\tan^2\theta(\Psi_0+\bar\Psi_0)$ and $G_\Phi$ all emerge as outputs. The equation $E_Z$ is not needed.

\paragraph{A boost-weight-zero transport identity.} Let $F=(n+3)r^2+(n-1)a^2x^2$, let $N$ be as in Sec.~\ref{sec:psi0deriv} and $Y=X_1/(3\Psi_2)$. The following combination of the identities is gauge invariant:
$$
\frac{n\bar\zeta^2}{2(n+2)r^2}({\rm e})+\frac{n\zeta^2}{2(n+2)r^2}(\bar{\rm e})-\frac{2\Sigma}{(n+2)r^2}({\rm CB})+c_E(E)+\bar c_E(\bar E)+\frac{12}{n+2}({\rm DL}),
$$
where
$$
c_E=-\frac{{\rm i}\zeta\,\big[(n-1)(n+2)a^3x^3-{\rm i}(n^2-1)a^2rx^2+n(n+3)ar^2x-{\rm i}(n-1)(n+3)r^3\big]}{(n+2)r^2F}.
$$
Setting this combination to zero and writing it out gives
\begin{equation}
DG_\Phi=A\,\bar\delta Y+\bar A\,\delta\bar Y+B\,Y+\bar B\,\bar Y+C\,\boldsymbol\Delta\Phi_{00}+\mathcal E\,\Phi_{00}+\mathcal F\,G_\Phi,\tag{R0}
\end{equation}
\begin{align*}
A&=\frac{\Sigma N}{2(n+2)r^2F},\qquad C=\frac{\Sigma\,[n(n+3)r^2-(n+2)(n-1)a^2x^2]}{(n+2)r^2F},\qquad \mathcal F=-\frac{(n+2)(n+3)r^2+(n+4)(n-1)a^2x^2}{rF},\\
B&=-\frac{\sqrt2\,{\rm i}}{4(n+2)r^2x\sin\theta\,F}\Big\{n(n-1)a^3x^3[(n+3)x^2-(n+2)]+{\rm i}(n-1)(n+6)a^2rx^2[(n+2)x^2-(n+1)]\\
&\hspace{4.2cm}+(n-4)(n+3)ar^2x[(n+1)x^2-n]+{\rm i}(n+2)(n+3)r^3[nx^2-(n-1)]\Big\},\\
\mathcal E&=\frac{1}{2(n+2)r^3F}\Big\{n(n+2)(n-1)a^4x^2+(n-1)(n+2)(n+4)a^2r^2x^2-n(n-2)(n+3)a^2r^2\\
&\hspace{2.6cm}-n(n+2)(n+3)r^4+q\big[(n-1)(n-2)(n+2)a^2x^2-n^2(n+3)r^2\big]\Big\},\qquad q=r_0^{n+1}r^{1-n}.
\end{align*}
(R0) is the boost-weight-zero partner of (R1) and (R1$'$). The coefficients are independent of $\omega$ and $m$.

\paragraph{The $\Phi_{00}$ component of Eq.~(\ref{eq:vecII}).} This step mirrors Sec.~\ref{sec:psi0deriv}. Write (R0) as $\mathcal G=A\bar\delta Y+\bar A\delta\bar Y+BY+\bar B\bar Y$, with $\mathcal G\equiv DG_\Phi-\mathcal FG_\Phi-C\boldsymbol\Delta\Phi_{00}-\mathcal E\Phi_{00}$. Write (R1) as $DY+a_1Y=S_1$. Using $[\bar\delta,D]=(\alpha+\bar\beta-\pi)D-\rho\bar\delta$, the combination
$$
(D-\lambda)({\rm R0})+A\,\bar\delta({\rm R1})+\bar A\,\delta(\bar{\rm R}1)+\big[B-A(\alpha+\bar\beta-\pi)\big]({\rm R1})+\big[\bar B-\bar A(\bar\alpha+\beta-\bar\pi)\big](\bar{\rm R}1),\qquad \lambda=D\ln A+\rho-a_1,
$$
is free of $Y$ and $\bar Y$. The $\delta\bar Y$, $Y$ and $\bar Y$ coefficients vanish identically on the background, just as $c_Y=0$ did in Sec.~\ref{sec:psi0deriv}. It equals
\begin{equation}
\Big\{r^{-(n+6)}D\big(r^{n+6}DG_\Phi\big)+\tfrac{2(n+3)}{r^2}G_\Phi-\tfrac{(K_3\Omega)_{\Phi_{00}}}{2r^2}-\tfrac{{\rm i}n[\omega(r^2-a^2)+am]}{2r^3}\Phi_{00}\Big\}+\kappa\,\Big[\Phi_{00}\ \text{component of (\ref{eq:vecI})}\Big],
\end{equation}
with $\kappa=C/(2\Sigma)=\frac{n(n+3)r^2-(n+2)(n-1)a^2x^2}{2(n+2)r^2F}$. On shell the left-hand side vanishes, which proves the $\Phi_{00}$ component of Eq.~(\ref{eq:vecII}).

In this step the $\Phi_{00}$ row of $K_3$ is again an output. Its operators $\mathcal L_1\mathcal L_2$ on $\Psi_0$ arise from $A\,\bar\delta(\bar\delta\Psi_0)$ inside $A\,\bar\delta({\rm R1})$, and the coupling ${\rm i}n[\omega(r^2-a^2)+am]/(2r^3)$ arises from the mode form of $D$. Here the trace equation $E_Z$ enters, through (DL).

\paragraph{Summary.} Both components of Eqs.~(\ref{eq:vecI})--(\ref{eq:vecII}) are derived from the Bianchi, Ricci, Codazzi and field equations.

\section{The physical subspace of the radial system}\label{sec:phys}
\paragraph{Notation.} We write $\beta=a\omega-m$, $X=\omega(r^2-a^2)+am$ and $q=r_0^{n+1}r^{1-n}$. Thus $\Delta=r^2+a^2-q$ and $2\Delta-r\Delta'=2a^2-(n+1)q$. Expanded, Eq.~(\lettereq{eq:radI}) reads $\Delta f''+c_1f'+(c_0+\lambda)f+2(n+2)g=0$, with
$$
c_1=3\Delta'+\frac{n\Delta}{r},\qquad c_0=V+\frac{3n\Delta'}{2r}+\frac{n(n-2)\Delta}{4r^2}-\frac{n(n+6)}4,
$$
where $V$ is the potential of Eq.~(\ref{eq:Dr}) [the part of $\mathcal D_r$ without derivatives]. Eq.~(\lettereq{eq:radII}) reads $g''+e_1g'+e_0g=s f$, with $e_1=\frac{n+6}r-\frac{2{\rm i}K}\Delta$, $e_0=-k'-\frac{n+6}rk+k^2+\frac{2(n+3)}{r^2}$, $k=\frac{{\rm i}K}\Delta$, and $s=\frac{\nu}{2r^2}+\frac{{\rm i}nX}{2r^3}$. Two polynomials recur:
\begin{equation}\label{eq:JT}
\begin{aligned}
\mathcal{J}&=4r^2K+2{\rm i}r\big[2a^2-(n+1)q\big],\\
T&=16rKX+{\rm i}\big[16\omega r^4+16a(2a\omega+\beta)r^2-8(n+3)\omega qr^2+8(n-1)a\beta q\big].
\end{aligned}
\end{equation}

\paragraph{The physical relations.} On physical solutions, the radial functions obey two first-order relations:
\begin{align}
&(\nu+\chi)\,f'+P_f\,f+P_g\,g=0,\label{eq:P1}\\
&(\nu+\chi)\,g'+Q_f\,f+\big[R_K(\nu+\chi)-\chi'-P_f\big]\,g=0.\label{eq:P2}
\end{align}
Their coefficients are
\begin{subequations}\label{eq:radcoeffs}
\begin{gather}
\chi=-\frac{{\rm i}n}{4}\frac{T}{\mathcal{J}},\qquad R_K=\frac{n+4}{r}+\frac{2\Delta'}{\Delta}-\frac{\mathcal{J}'}{\mathcal{J}},\label{eq:chiRK}\\
Q_f=-\frac{{\rm i}\Delta}{r^2\mathcal{J}}\big(n^2X^2+n\Delta\,\nu+r^2\nu^2\big),\\
P_f=\frac{1}{\Delta \mathcal{J}}\Big[\frac{p_0}{r}+p_1\nu+2r\Delta\lambda\,(nX+{\rm i}r\nu)\Big],\qquad
P_g=\frac{8(n+2){\rm i}r^4}{n\Delta \mathcal{J}}\,\Pi,\\
\Pi=\nu^2+\frac{2\lambda\nu}{n+2}-\frac{n\lambda^2}{2(n+2)}+\frac{n(n+3+\beta^2)}{n+2}\lambda+(C_1-f_1)\nu+C_0-f_0 .
\end{gather}
\end{subequations}
The quantities in $\Pi$ are
\begin{subequations}\label{eq:Piaux}
\begin{align}
f_1&={\rm i}n\Big(\omega r-\frac{a\beta}{r}\Big)+\frac{n(a^2-q)}{2r^2},\\
f_0&=\frac{n^2}{2}\Big(\omega^2r^2+\frac{a^2\beta^2}{r^2}\Big)+{\rm i}n^2\Big[\omega r-\frac{a(2a\omega+\beta)}{r}\Big]+\frac{{\rm i}n(n+3)\omega q}{2r}-\frac{{\rm i}n^2(n-1)a\beta q}{2(n+2)r^3},\\
C_1&=\frac{3n^2+2n-8-4\beta^2}{2(n+2)},\\
C_0&=-\frac{n}{2(n+2)}\Big[\beta^4+2n^2a\omega\beta+2(n+3)\beta(2a\omega+\beta)+4(n+2)\Big].
\end{align}
\end{subequations}
The polynomials $p_1$ and $p_0$ are
\begin{subequations}\label{eq:polynomials}
\begin{align}
p_1={}&-2{\rm i}(n^2-2)q^2+2q\big[(4n+3)\omega r^3+(4n-3)a\beta r-{\rm i}(n^2+3n+2-\beta^2)r^2-{\rm i}(n^2-3n+4)a^2\big]\nonumber\\
&+4{\rm i}\omega^2r^6+2(n+2)\omega r^5+2{\rm i}(4a\omega\beta-\beta^2-n-2)r^4+2a\big[(n-8)a\omega+(n+8)\beta\big]r^3\nonumber\\
&+2{\rm i}(\beta^2+6)a^2r^2+2(n-2)a^3\beta r+2{\rm i}na^4,\\
\tfrac{1}{2n}p_0={}&q^2\big[(n-1)(n-2)a\beta-n(n+3)\omega r^2\big]\nonumber\\
&+q\Big\{n(n-1)a^3\beta+{\rm i}(4n-5)a^2\beta^2r+2{\rm i}a\beta\omega r^3-ar^2\big[(n^2-5n-4)a\omega+\beta^3-(n^2+5n-10)\beta\big]\nonumber\\
&\hspace{2em}-{\rm i}(4n+5)\omega^2r^5-(n^2+n+2-\beta^2)\omega r^4\Big\}\nonumber\\
&+2\omega^3r^8-{\rm i}n\omega^2r^7+\beta\omega(2a\omega-\beta)r^6-{\rm i}a\omega\big[(n-10)a\omega+2\beta\big]r^5+a\big[\beta^3-3a\omega\beta^2+2(n+1)a\omega+2(n+5)\beta\big]r^4\nonumber\\
&-{\rm i}a^2\beta\big[2a\omega-(n+10)\beta\big]r^3+a^3\big[2(n-3)a\omega-\beta^3+2(n+1)\beta\big]r^2+{\rm i}na^4\beta^2r .
\end{align}
\end{subequations}
As PDE identities, Eqs.~(\ref{eq:P1})--(\ref{eq:P2}) hold with $\nu\to K_3$ and $\lambda\to K_C$ (in this order in the product $\lambda\nu$), acting on $\Omega$ and $\bm G$.

\paragraph{Closure.} Equations~(\ref{eq:P1})--(\ref{eq:P2}) are derived from the NP identities in Sec.~\ref{sec:physorigin}. Their compatibility with Eqs.~(\lettereq{eq:radI})--(\lettereq{eq:radII}) gives an independent check, and it also shows how rigid they are. Take the structure of Eqs.~(\ref{eq:P1})--(\ref{eq:P2}) and their simple $\lambda$-dependent coefficients as given. Then everything else follows from closure. Write the relations as $Y'=\mathsf MY$ for $Y=(f,g)$, and Eqs.~(\lettereq{eq:radI})--(\lettereq{eq:radII}) as $Y''=\mathsf AY'+\mathsf BY$. Consistency then requires $\mathsf M'+\mathsf M^2=\mathsf A\mathsf M+\mathsf B$ on the joint spectrum. Its four entries give the following.
\begin{itemize}
\item The closure condition multiplying $f$ in Eq.~(\lettereq{eq:radII}) holds identically. In the condition multiplying $g$ in Eq.~(\lettereq{eq:radI}), $P_f$ appears only through the sum $P_f+Q_g$, so $Q_g=R_O+R_K\nu-P_f$ with $R_O=R_K\chi-\chi'$. The condition is then linear in $P_g$. Writing $P_g=p_2\Pi$ with $p_2=8(n+2){\rm i}r^4/(n\Delta \mathcal{J})$, it reduces to $\partial_r\Pi=-\frac{{\rm i}n\,\mathcal{J}}{4r^4}(\nu+\chi)$. This integrates in closed form: $\partial_rf_1=\frac{{\rm i}n\,\mathcal{J}}{4r^4}$ and $\partial_rf_0=\frac{{\rm i}n\,\mathcal{J}\chi}{4r^4}$. The integration constants $C_1$ and $C_0$ are left undetermined at this stage.
\item The difference between the closure condition multiplying $f$ in Eq.~(\lettereq{eq:radI}), divided by $\Delta$, and that multiplying $g$ in Eq.~(\lettereq{eq:radII}) is linear in $P_f$. It holds identically in $(\lambda,\nu)$ and gives two first-order equations, $-2y'+u\,y=\text{source}$, for $y=\partial_\nu P_f$ and $y=P_f|_{\nu=\lambda=0}$, with $u=\frac2r+\frac{\Delta'}\Delta-\frac{2\mathcal{J}'}{\mathcal{J}}+\frac{2{\rm i}K}{\Delta}$. The homogeneous solution is $r\Delta^{1/2}\mathcal{J}^{-1}e^{{\rm i}\int K/\Delta}$, which is not rational. The rational solutions are therefore unique, and they are $p_1/(\Delta \mathcal{J})$ and $p_0/(r\Delta \mathcal{J})$.
\item The remaining condition, the closure condition multiplying $f$ in Eq.~(\lettereq{eq:radI}), holds only on the joint spectrum. Its numerator is a quartic polynomial in $(\lambda,\nu)$ with $r$-dependent coefficients. Requiring that it equal $(g_0+g_\nu\nu)F(\lambda,\nu)$, with $F$ independent of $r$, fixes $C_0$, $C_1$ and $F$ uniquely.
\end{itemize}
The resulting spectral curve, normalised so that the coefficient of $\lambda^2\nu$ is $1$, is
\begin{multline}
F=\lambda^2\nu-\frac2n\lambda\nu^2-\frac{n+2}{n}\nu^3-2(\beta^2+n+3)\lambda\nu+\frac2n\big[\beta^2-(n+2)(n-1)\big]\nu^2-8na\omega\beta\,\lambda\\
+\big[\beta^4+4n^2a\omega\beta+2n\beta(4a\omega+\beta)+6\beta(2a\omega+\beta)+4(n+2)\big]\nu+4na\omega\big[2\beta^3+(n-1)(n+3)a\omega+(n^2+4n+7)\beta\big].\label{eq:Fcurve}
\end{multline}
For $n=1,2,3$ and $a\omega=\frac12$, $m=1$ it reproduces the curves obtained from the angular operators in Sec.~\ref{sec:operators}. At $a=0$ it factorises into three lines for every $m$:
$$
F\big|_{a=0}=\frac{\nu}{n}\,\big(\lambda+\nu-m^2-2\big)\big[n\lambda-(n+2)\nu-n(m^2+2n+4)\big].
$$
These are the vector, tensor and scalar KI branches; for $m=2$ they reproduce Eq.~(\lettereq{eq:a0}). The angular spectral problem is thus encoded in the radial closure.

\paragraph{Second-order equation.} Eliminating $g$ with Eq.~(\ref{eq:P1}) gives
\begin{equation}
f''+\Big[\frac{3\Delta'}{\Delta}+\frac nr+\frac{{\rm i}n\,(4\mathcal{J}\,\nu-{\rm i}nT)}{16r^4\,\Pi}\Big]f'+\Bigg\{\frac{c_0+\lambda}{\Delta}+\frac{{\rm i}n\,\big[p_0/r+p_1\nu+2r\Delta\lambda(nX+{\rm i}r\nu)\big]}{4r^4\Delta\,\Pi}\Bigg\}f=0.\label{eq:phys2}
\end{equation}
The only singularities beyond those of Eq.~(\lettereq{eq:radI}) are at the zeros of $\Pi$, which are apparent. Then $g$ follows algebraically from Eq.~(\ref{eq:P1}). The pair $(f,g)$ obeys Eqs.~(\lettereq{eq:radI})--(\lettereq{eq:radII}) whenever $(\lambda,\nu)$ lies on the curve $F=0$.

\section{Derivation of the physical relations from the NP identities}\label{sec:physorigin}
This section derives Eqs.~(\ref{eq:P1})--(\ref{eq:P2}) and the curve $F=0$ from the NP identities. We use the Geroch--Held--Penrose (GHP) operators~\cite{Geroch:1973am} $\thorn,\thorn',\eth,\eth'$ on quantities of type $(p,q)$. In the Kinnersley frame $\thorn=D$, $\thorn'=\boldsymbol\Delta-p\gamma-q\bar\gamma$, $\eth=\delta-p\beta-q\bar\alpha$ and $\eth'=\bar\delta-p\alpha-q\bar\beta$, with $\rho'=-\mu$ and $\tau'=-\pi$. The bar denotes the $\bm m\leftrightarrow\bar{\bm m}$ partner. Write $Z=r\cos\theta$ for the warp factor, and
$$
k=\frac{n\bar\zeta F}{\zeta N},\qquad c_3=\frac{{\rm i}\zeta^2[(n-1)ax+{\rm i}(n+3)r]}{4anrx(ax+{\rm i}r)},\qquad x=\cos\theta .
$$

\paragraph{1. Gauge-invariant variables.} The first-order invariants used are:
\begin{itemize}
\item the boost-weight-$2$ components $\Psi_0$, $\Phi_{00}$ and their bars, and the boost-weight-zero components $\Phi_{02}$ and $\Phi_{20}$;
\item $Y=k\Psi_1-\Phi_{01}=X_1/(3\Psi_2)$ and its prime $Y'=k\Psi_3-\Phi_{21}$;
\item $H_1\equiv\psi_2+2c_3\varphi$, with $\psi_2$, $\varphi$ defined below, and its bar. For $n>1$, $H_1=\delta\Psi_2+c_3\,\delta R_{\bm m\bar{\bm m}}+c_4\,\delta R$, with
$$
c_4=\frac{(n-1)na^3x^3+{\rm i}(2n^2+n-3)a^2rx^2-(n^2+5n-6)ar^2x-3{\rm i}(n+3)r^3}{12an(n-1)x\,(ax+{\rm i}r)^2},
$$
the combination of $\delta\Psi_2$, $\delta R_{\bm m\bar{\bm m}}$ and $\delta R$ whose background gradient vanishes. The pole of $c_4$ at $n=1$ is spurious: on shell $\delta R=-n(n-1)\delta U$, and $n(n-1)c_4$ is regular.
\item the two warp-corrected invariants $\varphi$ and $\psi_2$.
\end{itemize}
Let $U\equiv[(\nabla Z)^2-1]/Z^2$. On the background $U=-r_0^{n+1}r^{-n-1}/\Sigma$ for every $n$, and $(Z,U)$ are local coordinates on the $(r,\theta)$ quotient away from the rotation axis: the Jacobian $\partial(Z,U)/\partial(r,\theta)$ is nonzero there. Every background scalar is therefore a function of $(Z,U)$; in particular $\Phi_{11}=f(Z,U)$ and $\Psi_2=h(Z,U)$. The perturbations
\begin{equation}
\varphi=\delta\Phi_{11}-\partial_Zf\,\delta Z-\partial_Uf\,\delta U,\qquad \psi_2=\delta\Psi_2-\partial_Zh\,\delta Z-\partial_Uh\,\delta U
\end{equation}
are therefore invariant under diffeomorphisms. To first order they are also invariant under tetrad rotations, since $\Psi_1=\Psi_3=0$ on the background and both have boost weight zero. For $n>1$ the trace of Eq.~(\lettereq{eq:reduced}) gives $\Lambda=R/24=-n(n-1)U/24$, so $(Z,U)$ may be replaced by $(Z,\Lambda)$ without changing $\varphi$ or $\psi_2$; this was the form of the original derivation. For $n=1$, $R=0$ on shell and only the $(Z,U)$ form is available. For $n>1$, $\psi_2+2c_3\varphi=\delta\Psi_2+c_3\,\delta R_{\bm m\bar{\bm m}}+c_4\,\delta R$. Since $\bar\zeta^2c_3+\zeta^2\bar c_3=-2\Sigma/n$,
\begin{equation}\label{eq:GPpsi}
2(n+2)r^2G_\Phi=n\big(\bar\zeta^2\psi_2+\zeta^2\bar\psi_2\big)-4\Sigma\,\varphi .
\end{equation}
This makes the gauge invariance of $G_\Phi$ manifest.

\paragraph{2. First-order identities.} Besides the NP identities, we use the trace relation in derivative form,
\begin{equation}
e(U)=\frac{2}{nZ}R(e,\nabla Z)-2U\frac{e(Z)}{Z}\qquad\text{for every tetrad vector }e,\tag{T}
\end{equation}
which follows from the definition of $U$, $e\big[(\nabla Z)^2\big]=2H(e,\nabla Z)$ and $H_{ab}=\frac ZnR_{ab}$. For $n>1$, multiplying (T) along $\bm l$ by $-n(n-1)/24$ gives (DL) of Sec.~\ref{sec:phi00deriv}. The identities are derived in a gauge where $\delta Z=\delta U=0$, $\Psi_1=\Psi_3=0$ and $\dot\epsilon=0$ (for $n>1$, $\delta U=0$ is $\delta\Lambda=0$). Explicit combinations of a few linearised NP identities then remove every perturbed spin coefficient and every component of the perturbed tetrad. Each result involves gauge invariants only and so holds in every gauge.
\begin{itemize}
\item The transport equations (R1) and (R1$'$) in Sec.~\ref{sec:psi0deriv} follow from the Bianchi identities (a), (b) and the Codazzi identities (C1), (C2).
\item The gauge-invariant form of the Bianchi identity (e). It uses (e), (CB1), the Codazzi components $(\bm l;\bm m,\bar{\bm m})$, $(\bm l;\bar{\bm m},\bm m)$ and (T) along $\bm l$. With $\beta_k\equiv1+\frac1k+2c_3$ and $E=-\frac{{\rm i}r}{anx}\big[(n-1)+\frac{4r^2}{\bar\zeta^2}\big]$:
\begin{multline}
\thorn H_1=\tfrac1k\,\eth'Y+\big(\tfrac1k+2c_3\big)\thorn'\Phi_{00}-\Big[2\bar\tau\big(\tfrac1k+2c_3\big)-2c_3\tau'+\big(2c_3-\tfrac n2\beta_k+E\big)\tfrac{\eth'Z}{Z}\Big]Y\\
-\Big[2\tau\big(\tfrac1k+2c_3\big)-2c_3\bar\tau'+(\beta_k-nc_3+E)\tfrac{\eth Z}{Z}\Big]\bar Y-\Big[\tfrac{\bar\rho'}k+\big(E-nc_3-\tfrac n2\beta_k\big)\tfrac{\thorn'Z}{Z}\Big]\Phi_{00}\\
+\Big(3\rho-\tfrac n2\beta_k\tfrac{\thorn Z}{Z}\Big)H_1-nc_3\tfrac{\thorn Z}{Z}\bar H_1 .
\tag{\(R_H\)}\label{eq:identityRH}
\end{multline}
\item The gauge-invariant form of the Bianchi identity (d). It uses (d), (CB2), the Codazzi components $(\bm l;\bm n,\bm m)$, $(\bm m;\bm m,\bar{\bm m})$ and (T) along $\bm m$:
\begin{multline}
\eth H_1=\tfrac1k\thorn'Y+\big(\tfrac1k+2c_3\big)\eth'\Phi_{02}+\Big\{-\tfrac{2\bar\rho'}k+2c_3(\rho'-2\bar\rho')+\big[E-\tfrac n2(1-\tfrac1k)-nc_3\big]\tfrac{\thorn'Z}Z\Big\}Y\\
+\Big\{-\tfrac{\bar\tau}k+\big[E-\tfrac n2(1-\tfrac1k)+2c_3\big]\tfrac{\eth'Z}Z\Big\}\Phi_{02}-\tfrac{nc_3}{\bar k}\tfrac{\thorn Z}Z\,\bar Y'
+\Big[3\tau-\tfrac n2\big(1-\tfrac1k-2c_3\big)\tfrac{\eth Z}Z\Big]H_1+nc_3\tfrac{\eth Z}Z\bar H_1 .
\tag{GI-d}\label{eq:identityGId}
\end{multline}
\item Two further identities each follow from a single Codazzi component and (T). Let
$$
E_\varphi=\frac{n-1}2+\frac{F^2-16a^2r^2x^2}{2\Sigma F},\qquad E'_\varphi=\frac{n-1}2-\frac{F^2-16a^2r^2x^2}{2\Sigma F},
$$
i.e.\ $E_\varphi=\frac{n-1}2+\frac4n\partial_Uf$ and $E'_\varphi=\frac{n-1}2-\frac4n\partial_Uf$, with $\partial_Uf=\frac{n(F^2-16a^2r^2x^2)}{8\Sigma F}$. (In the $\Lambda$ form, $E_\varphi=\frac{n-1}6(3-f_\Lambda)$ with $f_\Lambda=\partial_\Lambda f$; the factor $1/(n-1)$ in $f_\Lambda$ cancels.) The Codazzi component $(\bm l;\bm l,\bm n)$ plus a multiple of (T) along $\bm l$ gives
\begin{multline}
\thorn\varphi=\thorn'\Phi_{00}+(1+E_\varphi)\tfrac{\thorn'Z}{Z}\Phi_{00}+\big[\tau'-2\bar\tau+(E_\varphi-\tfrac n2)\tfrac{\eth'Z}{Z}\big]Y+\big[\bar\tau'-2\tau+(E_\varphi-\tfrac n2)\tfrac{\eth Z}{Z}\big]\bar Y\\
-(1+n-E_\varphi)\tfrac{\thorn Z}{Z}\varphi-\tfrac n2\tfrac{\thorn Z}{Z}(\psi_2+\bar\psi_2).
\tag{\(\Phi\text{-}\thorn\)}\label{eq:identityPhithorn}
\end{multline}
The component $(\bm m;\bm m,\bar{\bm m})$ plus a multiple of (T) along $\bm m$ gives
\begin{equation}
\eth\varphi=\eth'\Phi_{02}+(1+E'_\varphi)\tfrac{\eth'Z}{Z}\Phi_{02}+\big[\rho'-2\bar\rho'+(E'_\varphi-\tfrac n2)\tfrac{\thorn'Z}{Z}\big]Y-\tfrac{n}{2\bar k}\tfrac{\thorn Z}{Z}\,\bar Y'-(1+n-E'_\varphi)\tfrac{\eth Z}{Z}\varphi+\tfrac n2\tfrac{\eth Z}{Z}(\psi_2+\bar\psi_2).
\tag{\(\Phi\text{-}\eth\)}\label{eq:identityPhieth}
\end{equation}
\end{itemize}
No Ricci identity is used.

\paragraph{3. A closed transport pair for $\psi_2$.} Two combinations matter: (R$_H$)$-2c_3$($\Phi$-$\thorn$), an identity for $\thorn\psi_2$, and (GI-d)$-2c_3$($\Phi$-$\eth$), an identity for $\eth\psi_2$. In the second, $\bar Y'$ cancels identically, since its two coefficients differ by the factor $2c_3$. Using Eq.~(\ref{eq:GPpsi}) to eliminate $\varphi$, the result is
\begin{equation}\label{eq:psipairS}
\thorn\psi_2=\big(3\rho-\tfrac n2\tfrac{\thorn Z}Z\big)\psi_2+\mathfrak g\tfrac{\thorn Z}{Z}\bar\psi_2+\mathcal S_\thorn,\qquad
\eth\psi_2=\big(3\tau-\tfrac n2\tfrac{\eth Z}Z\big)\psi_2-\mathfrak g\tfrac{\eth Z}{Z}\bar\psi_2+\mathcal S_\eth,\qquad \mathfrak g=\frac{\zeta^3N}{2\bar\zeta^3F},
\end{equation}
with sources
\begin{align*}
\mathcal S_\thorn&=\tfrac1k(\eth'Y+\thorn'\Phi_{00})+\Big[\tfrac{2}{k}(\tau'-\bar\tau)+\big(1+\tfrac n2\big)\big(1+\tfrac1k\big)\tfrac{\eth'Z}Z\Big]Y+\Big[\tfrac{2\tau}k\tfrac{\thorn'Z}{\eth Z}-\tfrac{\bar\rho'}k+\big(1+\tfrac n2\big)\big(1+\tfrac1k\big)\tfrac{\thorn'Z}Z\Big]\Phi_{00}-\tfrac{\mathfrak g}{\zeta^2}\tfrac{\thorn Z}{Z}\mathcal G,\\
\mathcal S_\eth&=\tfrac1k(\thorn'Y+\eth'\Phi_{02})+\Big[\tfrac{2}{k}(\rho'-\bar\rho')+\big(1+\tfrac n2\big)\big(\tfrac1k-1\big)\tfrac{\thorn'Z}Z\Big]Y+\Big[\tfrac{2\rho}k\tfrac{\eth'Z}{\thorn Z}-\tfrac{\bar\tau}k+\big(1+\tfrac n2\big)\big(\tfrac1k-1\big)\tfrac{\eth'Z}Z\Big]\Phi_{02}+\tfrac{\mathfrak g}{\zeta^2}\tfrac{\eth Z}{Z}\mathcal G,
\end{align*}
and $\mathcal G=2(n+2)r^2G_\Phi/n$. Neither $Y'$ nor $\bar Y'$ appears. The pair is closed: its integrability condition on a mode is algebraic. With $\thorn=\partial_r-{\rm i}K/\Delta$ and $\eth=(\partial_\theta-Q)/(\sqrt2\zeta)$, the background part of the $(r,\theta)$ connection is flat, because $\mathfrak g$ is homogeneous of degree zero in $(r,\cos\theta)$. The only curvature is $Q$ times the commutator of the radial connection with $\mathrm{diag}(1,-1)$. Hence
$$
2Q\,\frac{\mathfrak g}{r}\,\bar\psi_2=\text{(local expression in }Y,\bar Y,\Phi_{00},\Phi_{02},\Phi_{20},G_\Phi\text{ and their first and second derivatives)},
$$
and likewise for $\psi_2$. In particular $\varphi$, $Y'$ and $\bar Y'$ are local on every mode with $Q\neq0$.

\paragraph{4. Reduction at the axis.} Equations~(\ref{eq:P1})--(\ref{eq:P2}) are obtained by evaluating the identities above at the rotation axis $\theta\to0$ on a regular separated mode. There, $\Omega=f(r)\bm Y(\theta)$ and $\bm G=g(r)\bm Y(\theta)$, with $K_C\bm Y=\lambda\bm Y$ and $K_3\bm Y=\nu\bm Y$, and $f,g$ obey Eqs.~(\lettereq{eq:radI})--(\lettereq{eq:radII}).
\begin{itemize}
\item \emph{Regularity.} A smooth perturbation has invariants of spin weight $s=(p-q)/2$ behaving as $\sin^{|m+s|}\theta$ times a smooth function of $\cos\theta$. For $m\ge2$ every invariant is $\sin^{m+s}\theta\,[\mathcal X_0(r)+\mathcal X_2(r)\sin^2\theta+\dots]$. The Frobenius coefficients of $\bm Y$ follow from $K_C$ and $K_3$. For example, normalising the leading coefficient of $Y_0$ to $1$, $Y_-\simeq4m(m-1)\sin^{m-2}\theta/(\nu+mn-2an\omega)$.
\item \emph{Operators.} Exactly,
$$
\eth\big(\sin^{m+s+j}\theta\big)=\tfrac{(m+j)x-m+a\omega\sin^2\theta}{\sqrt2\zeta}\sin^{m+s+j-1}\theta-q\bar\pi\sin^{m+s+j}\theta,
$$
$$
\eth'\big(\sin^{m+s+j}\theta\big)=\tfrac{(m+2s+j)x+m-a\omega\sin^2\theta}{\sqrt2\bar\zeta}\sin^{m+s+j-1}\theta-p\pi\sin^{m+s+j}\theta .
$$
So at leading order $\eth'$ multiplies by $2(m+s)/(\sqrt2\bar z)$, with $z=r+{\rm i}a$, while $\eth$ brings in the next coefficient. The radial operators act on the coefficients, and $f''$, $g''$ are eliminated with Eqs.~(\lettereq{eq:radI})--(\lettereq{eq:radII}).
\item \emph{Leading order.} The barred (R1$'$) is algebraic in $\bar Y_0$, and the barred (R1) then holds identically. The barred ($\Phi$-$\eth$) and (GI-d) are algebraic in $\varphi_0$, $\bar H_{1,0}$ in terms of $Y'_0$, and Eq.~(\ref{eq:GPpsi}) gives $H_{1,0}$. ($\Phi$-$\thorn$) is then a first-order equation for $Y'_0$, (R$_H$) gives $Y_0$, and (R1) fixes $Y'_0$. The single remaining compatibility condition is a multiple of $F(\lambda,\nu)$, so at leading order the identities reproduce the spectral curve and impose nothing else.
\item \emph{Next order.} The coefficients of relative order $\sin^2\theta$ are fixed in turn:
\begin{itemize}
\item[--] $\bar Y_2$ from the barred (R1$'$);
\item[--] $Y_2$ from (R1$'$) at leading order;
\item[--] $\psi_{2,2}$ and $\bar\psi_{2,2}$ from the barred ($\Phi$-$\eth$) and (GI-d), in terms of $Y'_2$;
\item[--] the leading coefficient of $\bar Y'$ from ($\Phi$-$\eth$);
\item[--] $Y'_2$ from (GI-d).
\end{itemize}
The remaining conditions of this order other than two hold identically on $F=0$. Those two, ($\Phi$-$\thorn$) and (R$_H$) at relative order $\sin^2\theta$, are first-order relations in $(f,f',g,g')$, and on $F=0$
$$
(\Phi\text{-}\thorn)_2=\alpha_1P_1+\beta_1P_2,\qquad ({\rm R}_H)_2=\alpha_2P_1+\beta_2P_2,\qquad \alpha_1\beta_2-\alpha_2\beta_1\neq0 .
$$
Hence Eqs.~(\ref{eq:P1})--(\ref{eq:P2}) hold on every regular mode with $m\ge2$.
\item \emph{Other $m$.} Complex conjugation maps a solution with $(\omega,m)$ to one with $(-\bar\omega,-m)$, exchanging barred and unbarred components. $P_1$, $P_2$ and $F$ are invariant under $(\omega,m,{\rm i})\to(-\omega,-m,-{\rm i})$, so $m\le-1$ follows from $m\ge1$. For $m=1$ the same reduction applies with leading powers $\sin^{|m+s|}\theta$. For $m=0$ the factor $2(m+s)$ vanishes on the spin-weight-zero invariants, and the reduction is carried to relative order $\sin^4\theta$. Thirty-two conditions (the identities at relative orders $0,2,4$ and some of their radial derivatives) leave, after elimination, one condition, $\alpha P_1+\beta P_2=0$ on $F=0$. The axis system at $m=0$ is invariant under the automorphism $\sigma$: ${\rm i}\to-{\rm i}$, $\omega\to-\omega$, combined with the exchange of barred and unbarred quantities (the algebraic form of complex conjugation), and $P_1$, $P_2$, $F$ are $\sigma$-invariant. Applying $\sigma$ to the identity gives $\sigma(\alpha)P_1+\sigma(\beta)P_2=0$ on the same mode, and $\alpha\sigma(\beta)-\beta\sigma(\alpha)\not\equiv0$ mod $F$. Hence Eqs.~(\ref{eq:P1})--(\ref{eq:P2}) hold for every $m$.
\end{itemize}
The steps after item~3 are linear algebra over the field of rational functions of $(r,q,n,a,\omega,m,\lambda,\nu)$ over $\mathbb Q({\rm i},\sqrt2)$, with $q=r_0^{n+1}r^{1-n}$ and $\partial_rq=(1-n)q/r$. They were carried out in exact arithmetic with all parameters symbolic ($m$ fixed for $m=0,1$). The statements ``multiple of $F$'' and ``lies in $\mathrm{span}(P_1,P_2)$ on $F=0$'' were decided by exact polynomial division by $F$.

\section{\texorpdfstring{The case $n=1$ ($d=5$)}{The five-dimensional case}}\label{sec:none}
For $n=1$ we consider the perturbations even under the reflection of the transverse circle (Sec.~\ref{sec:conventions}); the odd ones, with $h_{a\psi}\neq0$, form a decoupled sector not treated here. The derivation above is written for symbolic $n$, and three of its ingredients degenerate at $n=1$: the Ricci scalar vanishes on shell, $R=24\Lambda=-n(n-1)U=0$, so $\Lambda$ cannot serve as a background coordinate; $f_\Lambda$ and $c_4$ have simple poles at $n=1$; and (DL) becomes trivial. We therefore do not take a limit $n\to1$. Instead, every step is formulated so that it holds at $n=1$ as it stands, and is verified directly at $n=1$.
\begin{itemize}
\item \emph{Variables.} The warp-corrected invariants are defined through $(Z,U)$ (Sec.~\ref{sec:physorigin}.1), which are local background coordinates for every $n$ in the interior of the quotient; $H_1\equiv\psi_2+2c_3\varphi$; the gauge is $\delta Z=\delta U=0$. Eq.~(\ref{eq:GPpsi}) holds with $G_\Phi$ in the bracket form of Sec.~\ref{sec:phi00components}, for symbolic $n$ and at $n=1$. For $n>1$ all of these coincide with the $\Lambda$ forms.
\item \emph{Identities.} (DL) and the derivatives of the trace relation are replaced by (T), which is derived from the definition of $U$ and $H_{ab}=\frac ZnR_{ab}$ alone, both valid for $n=1$. The identities of Secs.~\ref{sec:psi0deriv} and \ref{sec:phi00deriv} use $\Lambda$ only as a curvature component, never as a coordinate; at $n=1$ they hold with $\delta\Lambda$ arbitrary, and in particular with $\delta\Lambda=0$.
\item \emph{Coefficients.} Apart from $f_\Lambda$ and $c_4$, which no longer appear, every coefficient used is regular at $n=1$: $k$, $c_3$, $\mathfrak g$, $E$, $E_\varphi$, $E'_\varphi$ and the coefficients of (R$_H$), (GI-d), ($\Phi$-$\thorn$), ($\Phi$-$\eth$); the combination coefficients of (R1), (R1$'$) in Sec.~\ref{sec:psi0deriv}, of (R0) and of $W$ in Sec.~\ref{sec:phi00deriv}; $F(\lambda,\nu)$; and the coefficients of Eqs.~(\ref{eq:P1})--(\ref{eq:P2}). The operators $K_C$, $K_3$ and Eqs.~(\lettereq{eq:radI})--(\lettereq{eq:radII}) are polynomial in $n$. Each rational coefficient was checked to have no denominator vanishing at $n=1$.
\item \emph{Direct verification at $n=1$.} With $n$ set to $1$ in the exact arithmetic (not as a limit), we re-verified: (C1), (C2) as Codazzi components and the combinations for (R1), (R1$'$); the generalised Teukolsky identity and the master equation (\ref{eq:master}); the $\Psi_0$ and $\Phi_{00}$ components of Eqs.~(\ref{eq:vecI})--(\ref{eq:vecII}), including (R0); Eq.~(\ref{eq:GPpsi}) and $H_1=\psi_2+2c_3\varphi$ in the $(Z,U)$ form; all eleven first-order identities of Sec.~\ref{sec:physorigin}.2 as combinations of linearised NP identities and (T) in the gauge $\delta Z=\delta U=0$, with the minimal identity sets quoted there; and the axis reduction for $m\ge2$ ($m$ symbolic), $m=1$ and $m=0$, including the $\sigma$-symmetry and the independence of the pair $\{\alpha P_1+\beta P_2,\ \sigma(\alpha)P_1+\sigma(\beta)P_2\}$ modulo $F$.
\end{itemize}
The analytic argument therefore covers $d=5$ (in the even sector) without a limiting step.

\section{Angular problem}\label{sec:angular}
In terms of $x=\cos\theta$ we set $Y_s=(1-x)^{|m+s|/2}u_s$ for $s=(+2,-2,0)$. After multiplication by $x^2$, the coefficients of $K_C$ and $K_3$ are regular on $[0,1]$. The regular solutions near $x=0$ are fixed by the $1/x^2$ part of $K_C$, which is $-\mathsf N/x^2$ with
\begin{equation}
\mathsf N=\begin{pmatrix} n&0&n+2\\0&n&n+2\\ \frac n2&\frac n2&n+2\end{pmatrix},\qquad \mathrm{spec}\,\mathsf N=\{0,n,2n+2\}.
\end{equation}
The corresponding regular exponents are $0,1,2$; the irregular ones are $1-n,-n,-(n+1)$, with a logarithm for $n=1$. Smoothness requires $\mathsf N\bm Y(0)=0$ and $\ell\cdot\bm Y'(0)=0$, with $\ell=(1,1,-2)$ satisfying $\ell\mathsf N=0$. In the factored variables, the derivative condition reads
$$
\ell\cdot\bm u'(0)=\frac{|m+2|}{2}u_{+2}(0)+\frac{|m-2|}{2}u_{-2}(0)-|m|u_0(0).
$$
In collocation, two endpoint rows impose $\mathsf N\bm u(0)=0$ (rank two). The third is obtained by differentiating $x^2K_C\bm Y=\lambda x^2\bm Y$ with respect to $x$ at $x=0$ and contracting with $\ell$. On the kernel of $\mathsf N$ this row equals $n\ell\cdot\bm Y'(0)$; evaluated in the factored variables, it includes the derivatives of the prefactors automatically. The condition at $x=1$ is natural. With 37 Chebyshev--Lobatto points the eigenvalues are converged to $10^{-10}$; at $a=0$, $\lambda$ is integral and $\nu$ takes the rational KI values to that accuracy.

At $a=0$ and $m=2$ we find Eq.~(\lettereq{eq:a0}) for $n=1,2,3,4$ (checked for the first seven eigenvalues in each case). For $d=6$ we have also confirmed the sector identification by computing quasinormal modes from each branch and comparing with Kodama--Ishibashi values. There is no $\tilde\ell_T=2$ member. This is consistent with the absence of an $SO(n+1)$-singlet with $SO(2)$ charge $2$ in the $\tilde\ell=2$ tensor representation.

\section{Radial problem and numerics}\label{sec:numerics}
We followed the numerical methods developed in Ref.~\cite{Dias:2015nua}. All numerical length scales are quoted in units of $r_+$.

\paragraph{Local structure.} We integrate the first-order system~(\ref{eq:P1})--(\ref{eq:P2}), written as $Y'=M(r)Y$ with $Y=(f,g)$. Its coefficients are rational in $r$ (with $q=r_0^{n+1}r^{1-n}$); besides $r=0$ and the roots of $\Delta$, its only singular points are the zeros of $\nu+\chi(r)$, which are apparent. At $r=r_+$, with $\epsilon=r-r_+$, $\epsilon M$ is analytic and $\epsilon M|_{\epsilon=0}$ has eigenvalues $-2-{\rm i}\kappa$ (ingoing) and ${\rm i}\kappa$ (outgoing), $\kappa=K(r_+)/\Delta'(r_+)$. We obtain the Taylor coefficients of $\epsilon M$ by a fast Fourier transform on a circle $|\epsilon|=\rho$, with $\rho$ at most $0.6$ times the distance from $r_+$ to the nearest other singular point (and $\rho\le0.3r_+$), compute the ingoing Frobenius series to 24 terms, and start the integration at $|\epsilon|=\rho/2$. At large $r$, for $\omega\ne0$, the coefficients of $f'$ and $f$ in Eq.~(\ref{eq:phys2}) behave as $(n+4)/r+O(r^{-2})$ and $\omega^2+2{\rm i}\omega/r+O(r^{-2})$, respectively. Substituting $f\sim e^{\pm{\rm i}\omega r}r^p$ gives $p=-(n+2)/2$ for the minus sign and $p=-(n+6)/2$ for the plus sign. The two solutions are therefore $f_{\rm in}\sim e^{-{\rm i}\omega r}r^{-(n+2)/2}$ and $f_{\rm out}\sim e^{+{\rm i}\omega r}r^{-(n+6)/2}$, ingoing and outgoing, respectively, with our $e^{-{\rm i}\omega t}$ convention.

\paragraph{Quasinormal condition.} Quasinormal modes are ingoing at the horizon and outgoing at infinity. We integrate the ingoing solution along $r=r_++s\,e^{{\rm i}\varphi}$ with $\varphi=\pi/2-\arg\omega$, up to $|s|=\max(30r_+,25/|\omega|)$, using an 8th-order Runge--Kutta method with relative tolerance $10^{-11}$, and form $A(\omega)=f\,e^{{\rm i}\omega r}r^{(n+2)/2}$ at the endpoint. Along this ray outgoing solutions are exponentially suppressed, so $A(\omega)=0$ sets the ingoing amplitude to zero and enforces outgoing behaviour at infinity. Its roots are found by the secant method, with the angular eigenvalues $(\lambda,\nu)(a\omega)$ tracked continuously.

\paragraph{Convergence and comparison with the fourth-order system.} Changing the angular resolution from 31 to 45 points, the ray angle by $\pm0.3$, the endpoint by a factor of $2$, and the integration tolerance to $10^{-13}$ changes $\omega$ by at most $3\times10^{-8}$. Integrating instead Eqs.~(\lettereq{eq:radI})--(\lettereq{eq:radII}), whose four-dimensional solution space at $r_+$ must be restricted to the ingoing exponent by monodromy, reproduces these frequencies to $10^{-7}$ on the $\mathrm{Re}\,\omega>0$ branches and to within $2\times10^{-4}$ on the $\mathrm{Re}\,\omega<0$ branches. The larger differences occur at large $a$ and come from the fourth-order computation: there its results move by up to $5\times10^{-4}$ when the ray angle is changed by $\pm0.3$, while those of the first-order system move by less than $10^{-7}$. The onset rotations in Table~\lettereq{tab:onset} agree between the two systems to $10^{-5}$.

Integrating Eqs.~(\lettereq{eq:radI})--(\lettereq{eq:radII}) numerically from data that satisfy Eqs.~(\ref{eq:P1})--(\ref{eq:P2}) preserves them to $10^{-12}$ and reproduces the solution of Eq.~(\ref{eq:phys2}). Generic data do not do so.

\paragraph{Ultraspinning modes.} For $m=0$ and $\omega={\rm i}\gamma$ the coefficients of Eqs.~(\ref{eq:P1})--(\ref{eq:P2}) are real on the real $r$ axis, and so are some of the apparent singular points. We therefore integrate along $r=r_++s\,e^{0.35{\rm i}}$, up to $|s|=\max(60r_+,15/\gamma)$, so that the outgoing solution, which decays as $e^{-\gamma r}$, is negligible at the endpoint. $A({\rm i}\gamma)$ then has a fixed phase, and its zeros are found by Brent's method. The onset $a_c$ follows from a quadratic fit of $\gamma(a)$ at three rotations with $\gamma r_+\lesssim3\times10^{-3}$; linear and quadratic fits agree to $10^{-4}$ in $a_c/r_+$. At $a/r_+=4.04$ in $d=6$ the fourth-order system gives the same $\gamma$ to $2\times10^{-7}$. As $\gamma\to0$ the angular eigenvalues approach the $a=0$ values $(2-C_{\tilde\ell},C_{\tilde\ell})$ of the KI tensor harmonics, as they must, since for $m=0$ the angular operators depend on $a$ only through $a\omega$.

The first ultraspinning branch is now followed to $a/r_+=8$ in each of $d=6,7,8,9$. Table~\ref{tab:us-extended} summarises the largest sampled growth rates and the values at the endpoint. The $d=7,8,9$ branches turn over within this range, while the $d=6$ branch remains increasing. All four modes remain unstable at the endpoint. The locations of the sampled maxima are listed without fitting between data points.

\begin{table}[ht]
\caption{Extended first-branch ultraspinning data ($m=0$, connected to $\tilde\ell_T=2$). $x_{\max}$ is the value of $a/r_+$ with the largest growth rate among the computed samples, rather than a fitted maximum. For $d=6$ this is the endpoint of the computed range.}
\label{tab:us-extended}
\begin{ruledtabular}
\begin{tabular}{lccc}
 & $x_{\max}$ & $(\gamma r_+)_{\max}$ & $\gamma r_+$ at $a/r_+=8$\\\hline
$d=6$ & 8.00 & 0.0728361 & 0.0728361\\
$d=7$ & 6.70 & 0.1011939 & 0.0985016\\
$d=8$ & 5.71 & 0.1138220 & 0.1034249\\
$d=9$ & 5.10 & 0.1207119 & 0.1033521\\
\end{tabular}
\end{ruledtabular}
\end{table}

\begin{table}[ht]
\caption{Quasinormal frequencies $\omega r_+$ of the $\ell_S=0$, $m=2$ modes connected to $\tilde\ell_S=2$.}
\begin{ruledtabular}
\begin{tabular}{lcccc}
$a/r_+$ & $0.5$ & $1.0$ & $1.5$ & $2.0$\\\hline
$d=5$ & $0.95251-0.16563{\rm i}$ & $0.82473-0.07852{\rm i}$ & $0.68544-0.03346{\rm i}$ & $0.57326-0.01384{\rm i}$\\
      & $-0.77616-0.25924{\rm i}$ & $-0.58825-0.21154{\rm i}$ & $-0.45296-0.16731{\rm i}$ & $-0.36192-0.13526{\rm i}$\\
$d=6$ & $1.02917-0.12877{\rm i}$ & $0.83418+0.02216{\rm i}$ & $0.66345+0.09641{\rm i}$ & $0.53632+0.12605{\rm i}$\\
      & $-1.04775-0.35884{\rm i}$ & $-0.88321-0.32953{\rm i}$ & $-0.74355-0.28673{\rm i}$ & $-0.63957-0.24997{\rm i}$\\
$d=7$ & $1.11597-0.14415{\rm i}$ & $0.87318+0.05001{\rm i}$ & $0.67859+0.13309{\rm i}$ & $0.53910+0.16086{\rm i}$\\
      & $-1.34406-0.47679{\rm i}$ & $-1.20600-0.45428{\rm i}$ & $-1.06921-0.41254{\rm i}$ & $-0.96140-0.37398{\rm i}$\\
\end{tabular}
\end{ruledtabular}
\end{table}

At $a=0$ in $d=6$, $(\lambda,\nu)=(-8,0)$, $(-16,-14)$ and $(-22,28)$ give $\omega r_+=2.188071-0.467877{\rm i}$, $2.623066-0.437897{\rm i}$ and $3.147805-0.497326{\rm i}$, respectively. These reproduce the KI modes $\tilde\ell_V=3$, $\tilde\ell_S=4$ and $\tilde\ell_T=4$ of Eq.~(\lettereq{eq:a0}).

\fi
\ifPRLsupplementonly
  \bibliographystyle{apsrev4-2}
  \bibliography{references}

@article{Teukolsky:1972my,
    author = "Teukolsky, S. A.",
    title = "{Rotating black holes - separable wave equations for gravitational and electromagnetic perturbations}",
    reportNumber = "OAP-291",
    doi = "10.1103/PhysRevLett.29.1114",
    journal = "Phys. Rev. Lett.",
    volume = "29",
    pages = "1114--1118",
    year = "1972"
}

@article{Teukolsky:1973ha,
    author = "Teukolsky, Saul A.",
    title = "{Perturbations of a rotating black hole. 1. Fundamental equations for gravitational electromagnetic and neutrino field perturbations}",
    doi = "10.1086/152444",
    journal = "Astrophys. J.",
    volume = "185",
    pages = "635--647",
    year = "1973"
}

@article{Durkee:2010qu,
    author = "Durkee, Mark and Reall, Harvey S.",
    title = "{Perturbations of higher-dimensional spacetimes}",
    eprint = "1009.0015",
    archivePrefix = "arXiv",
    primaryClass = "gr-qc",
    doi = "10.1088/0264-9381/28/3/035011",
    journal = "Class. Quant. Grav.",
    volume = "28",
    pages = "035011",
    year = "2011"
}

@article{Durkee:2010ea,
    author = "Durkee, Mark and Reall, Harvey S.",
    title = "{Perturbations of near-horizon geometries and instabilities of Myers-Perry black holes}",
    eprint = "1012.4805",
    archivePrefix = "arXiv",
    primaryClass = "hep-th",
    doi = "10.1103/PhysRevD.83.104044",
    journal = "Phys. Rev. D",
    volume = "83",
    pages = "104044",
    year = "2011"
}

@article{Myers:1986un,
    author = "Myers, Robert C. and Perry, M. J.",
    title = "{Black Holes in Higher Dimensional Space-Times}",
    reportNumber = "PRINT-86-0067 (PRINCETON)",
    doi = "10.1016/0003-4916(86)90186-7",
    journal = "Annals Phys.",
    volume = "172",
    pages = "304",
    year = "1986"
}

@article{Murata:2008yx,
    author = "Murata, Keiju and Soda, Jiro",
    title = "{Stability of Five-dimensional Myers-Perry Black Holes with Equal Angular Momenta}",
    eprint = "0803.1371",
    archivePrefix = "arXiv",
    primaryClass = "hep-th",
    reportNumber = "KUNS-2131, CAS-KITPC-021",
    doi = "10.1143/PTP.120.561",
    journal = "Prog. Theor. Phys.",
    volume = "120",
    pages = "561--579",
    year = "2008"
}

@article{Frolov:2017kze,
    author = "Frolov, Valeri P. and Krtous, Pavel and Kubiznak, David",
    title = "{Black holes, hidden symmetries, and complete integrability}",
    eprint = "1705.05482",
    archivePrefix = "arXiv",
    primaryClass = "gr-qc",
    doi = "10.1007/s41114-017-0009-9",
    journal = "Living Rev. Rel.",
    volume = "20",
    number = "1",
    pages = "6",
    year = "2017"
}

@article{Lunin:2017drx,
    author = "Lunin, Oleg",
    title = "{Maxwell{\textquoteright}s equations in the Myers-Perry geometry}",
    eprint = "1708.06766",
    archivePrefix = "arXiv",
    primaryClass = "hep-th",
    doi = "10.1007/JHEP12(2017)138",
    journal = "JHEP",
    volume = "2017",
    number = "12",
    pages = "138",
    year = "2017"
}

@article{Krtous:2018bvk,
    author = "Krtou{\v{s}}, Pavel and Frolov, Valeri P. and Kubiz{\v{n}}{\'a}k, David",
    title = "{Separation of Maxwell equations in Kerr{\textendash}NUT{\textendash}(A)dS spacetimes}",
    eprint = "1803.02485",
    archivePrefix = "arXiv",
    primaryClass = "hep-th",
    doi = "10.1016/j.nuclphysb.2018.06.019",
    journal = "Nucl. Phys. B",
    volume = "934",
    pages = "7--38",
    year = "2018"
}

@article{Emparan:2003sy,
    author = "Emparan, Roberto and Myers, Robert C.",
    title = "{Instability of ultra-spinning black holes}",
    eprint = "hep-th/0308056",
    archivePrefix = "arXiv",
    doi = "10.1088/1126-6708/2003/09/025",
    journal = "JHEP",
    volume = "2003",
    number = "09",
    pages = "025",
    year = "2003"
}

@article{Dias:2009iu,
    author = "Dias, Oscar J. C. and Figueras, Pau and Monteiro, Ricardo and Santos, Jorge E. and Emparan, Roberto",
    title = "{Instability and new phases of higher-dimensional rotating black holes}",
    eprint = "0907.2248",
    archivePrefix = "arXiv",
    primaryClass = "hep-th",
    doi = "10.1103/PhysRevD.80.111701",
    journal = "Phys. Rev. D",
    volume = "80",
    pages = "111701",
    year = "2009"
}

@article{Dias:2010eu,
    author = "Dias, Oscar J. C. and Figueras, Pau and Monteiro, Ricardo and Santos, Jorge E.",
    title = "{Ultraspinning instability of rotating black holes}",
    eprint = "1006.1904",
    archivePrefix = "arXiv",
    primaryClass = "hep-th",
    doi = "10.1103/PhysRevD.82.104025",
    journal = "Phys. Rev. D",
    volume = "82",
    pages = "104025",
    year = "2010"
}

@article{Dias:2010maa,
    author = "Dias, Oscar J. C. and Figueras, Pau and Monteiro, Ricardo and Reall, Harvey S. and Santos, Jorge E.",
    title = "{An instability of higher-dimensional rotating black holes}",
    eprint = "1001.4527",
    archivePrefix = "arXiv",
    primaryClass = "hep-th",
    doi = "10.1007/JHEP05(2010)076",
    journal = "JHEP",
    volume = "2010",
    number = "05",
    pages = "076",
    year = "2010"
}

@article{Figueras:2017zwa,
    author = "Figueras, Pau and Kunesch, Markus and Lehner, Luis and Tunyasuvunakool, Saran",
    title = "{End Point of the Ultraspinning Instability and Violation of Cosmic Censorship}",
    eprint = "1702.01755",
    archivePrefix = "arXiv",
    primaryClass = "hep-th",
    doi = "10.1103/PhysRevLett.118.151103",
    journal = "Phys. Rev. Lett.",
    volume = "118",
    number = "15",
    pages = "151103",
    year = "2017"
}

@article{Shibata:2010wz,
    author = "Shibata, Masaru and Yoshino, Hirotaka",
    title = "{Bar-mode instability of rapidly spinning black hole in higher dimensions: Numerical simulation in general relativity}",
    eprint = "1004.4970",
    archivePrefix = "arXiv",
    primaryClass = "gr-qc",
    doi = "10.1103/PhysRevD.81.104035",
    journal = "Phys. Rev. D",
    volume = "81",
    pages = "104035",
    year = "2010"
}

@article{Dias:2014eua,
    author = "Dias, {\'O}scar J. C. and Hartnett, Gavin S. and Santos, Jorge E.",
    title = "{Quasinormal modes of asymptotically flat rotating black holes}",
    eprint = "1402.7047",
    archivePrefix = "arXiv",
    primaryClass = "hep-th",
    doi = "10.1088/0264-9381/31/24/245011",
    journal = "Class. Quant. Grav.",
    volume = "31",
    number = "24",
    pages = "245011",
    year = "2014"
}

@article{Kodama:2003jz,
    author = "Kodama, Hideo and Ishibashi, Akihiro",
    title = "{A Master equation for gravitational perturbations of maximally symmetric black holes in higher dimensions}",
    eprint = "hep-th/0305147",
    archivePrefix = "arXiv",
    doi = "10.1143/PTP.110.701",
    journal = "Prog. Theor. Phys.",
    volume = "110",
    pages = "701--722",
    year = "2003"
}

@article{Lunin:2025gw,
    author = "Lunin, Oleg",
    title = "{Gravitational waves in the Myers-Perry geometry}",
    eprint = "2510.14417",
    archivePrefix = "arXiv",
    primaryClass = "hep-th",
    doi = "10.1007/JHEP03(2026)055",
    journal = "JHEP",
    volume = "2026",
    number = "03",
    pages = "055",
    year = "2026"
}

@misc{Hintz:2026kn,
    author = "Hintz, Peter",
    title = "{Perturbations of Kerr-Newman black holes: separation and mode stability}",
    eprint = "2609.33661",
    archivePrefix = "arXiv",
    primaryClass = "gr-qc",
    month = "9",
    year = "2026"
}

@misc{Destounis:2026kn,
    author = "Destounis, Kyriakos and Cardoso, Vitor and Hintz, Peter",
    title = "{The quasinormal modes of Kerr-Newman black holes: a separated approach}",
    eprint = "2610.07142",
    archivePrefix = "arXiv",
    primaryClass = "gr-qc",
    month = "10",
    year = "2026"
}

@article{Kodama:2009rq,
    author = "Kodama, Hideo and Konoplya, R. A. and Zhidenko, Alexander",
    title = "{Gravitational stability of simply rotating Myers-Perry black holes: Tensorial perturbations}",
    eprint = "0904.2154",
    archivePrefix = "arXiv",
    primaryClass = "gr-qc",
    doi = "10.1103/PhysRevD.81.044007",
    journal = "Phys. Rev. D",
    volume = "81",
    pages = "044007",
    year = "2010"
}

@article{Whiting:1988vc,
    author = "Whiting, Bernard F.",
    title = "{Mode Stability of the Kerr Black Hole}",
    reportNumber = "IFP-306-UNC",
    doi = "10.1063/1.528308",
    journal = "J. Math. Phys.",
    volume = "30",
    pages = "1301",
    year = "1989"
}

@FOOTNOTE{SupplementalMaterial,
  key = {SupplementalMaterial},
  note = {See Supplemental Material for the operator identities, the Newman--Penrose derivation, the physical radial relations, and the angular and radial numerical methods}
}

@CONTROL{PRLControl, title = "0"}

@article{Dias:2015nua,
    author = "Dias, {\'O}scar J. C. and Santos, Jorge E. and Way, Benson",
    title = "{Numerical Methods for Finding Stationary Gravitational Solutions}",
    eprint = "1510.02804",
    archivePrefix = "arXiv",
    primaryClass = "hep-th",
    doi = "10.1088/0264-9381/33/13/133001",
    journal = "Class. Quant. Grav.",
    volume = "33",
    number = "13",
    pages = "133001",
    year = "2016"
}

@article{Cardoso:2003qd,
    author = "Cardoso, Vitor and Lemos, Jose P. S. and Yoshida, Shijun",
    title = "{Scalar gravitational perturbations and quasinormal modes in the five-dimensional Schwarzschild black hole}",
    eprint = "hep-th/0311260",
    archivePrefix = "arXiv",
    reportNumber = "DF-IST-17-2003",
    doi = "10.1088/1126-6708/2003/12/041",
    journal = "JHEP",
    volume = "2003",
    number = "12",
    pages = "041",
    year = "2003"
}

@article{Yoshino:2005ps,
    author = "Yoshino, Hirotaka and Shiromizu, Tetsuya and Shibata, Masaru",
    title = "{The Close limit analysis for head-on collision of two black holes in higher dimensions: Brill-Lindquist initial data}",
    eprint = "gr-qc/0508063",
    archivePrefix = "arXiv",
    doi = "10.1103/PhysRevD.72.084020",
    journal = "Phys. Rev. D",
    volume = "72",
    pages = "084020",
    year = "2005"
}

@article{Geroch:1973am,
    author = "Geroch, Robert P. and Held, A. and Penrose, R.",
    title = "{A space-time calculus based on pairs of null directions}",
    doi = "10.1063/1.1666410",
    journal = "J. Math. Phys.",
    volume = "14",
    pages = "874--881",
    year = "1973"
}
\fi
\end{document}